\documentclass[aps,prb,twocolumn,showpacs,floatfix]{revtex4-1}
\usepackage{graphicx}
\usepackage{dcolumn}
\usepackage{bm}
\usepackage{natbib}
\usepackage{caption}
\usepackage{subcaption}
\usepackage{float}
\usepackage{amsmath,amssymb}
\usepackage{amsmath}
\usepackage{braket}
\usepackage{xcolor}
\usepackage[referable]{threeparttablex}

\usepackage[colorinlistoftodos, textsize=small]{todonotes}

\newcommand{\comment}[1]{}

\newcommand{\subscript}[1]{\ensuremath{_{\textrm{#1}}}}

\begin{document}
\title{103-Compound Band Structure Benchmark of Post-SCF Spin-Orbit Coupling Treatments in Density-Functional Theory}
\author{William P. Huhn}
\affiliation{Department of Mechanical Engineering and Materials Science, Duke University, Durham, NC, 27708}
\author{Volker Blum}
\affiliation{Department of Mechanical Engineering and Materials Science, Duke University, Durham, NC, 27708}
\begin{abstract}
We quantify the accuracy of different non-self-consistent and self-consistent spin-orbit coupling (SOC) treatments in Kohn-Sham and hybrid density-functional theory by providing a band structure benchmark set for the valence and low-lying conduction energy bands of 103 inorganic compounds, covering chemical elements up to Po. Reference energy band structures for the PBE density functional are obtained using the full-potential (linearized) augmented plane wave code Wien2k, employing its self-consistent treatment of SOC including Dirac-like $p^{1/2}$ orbitals in the basis set. We use this benchmark set to benchmark a computationally simpler, non-self-consistent all-electron treatment of SOC based on scalar-relativistic orbitals and numeric atom-centered orbital basis functions. For elements up to Z$\approx$50, both treatments agree virtually exactly. For the heaviest elements considered (Tl, Pb, Bi, Po), the band structure changes due to SOC are captured with a relative deviation of 11\% or less.  For different density functionals (PBE vs. the hybrid HSE06), we show that the effect of spin-orbit coupling is usually similar but can be dissimilar if the qualitative features of the predicted underlying scalar-relativistic band structures do not agree.  All band structures considered in this work are available online via the NOMAD Repository to aid in future benchmark studies and methods development.

\end{abstract}
\maketitle

\section{Introduction}

Spin-orbit coupling (SOC) is an essential ingredient for quantitatively correct energy band structures of materials composed of any but the few lightest elements, appearing in materials applications as diverse as heavy-light hole masses in conventional semiconductors,\cite{GDresselhaus54,GDresselhaus55,MSDresselhaus08} Rashba splittings in reduced dimensionality systems,\cite{DBercioux15,MKepenekian15,GBihlmayer15} topologically insulating phases,\cite{CLKane05} and Berry phase physics\cite{DXiao10}.  Yet, for reasons of cost and convenience, the effects of spin-orbit coupling in different computational studies are often approximated based on different underlying scalar-relativistic orbitals and, at the simplest level, in a non-self-consistent fashion.  In standard or generalized Kohn-Sham density-functional theory ((g)KSDFT),\cite{WKohn65,ABecke93_B3LYP,ASeidl96} the effects of SOC on electronic levels can be incorporated into calculations by way of the spin-orbit-coupled effective Hamiltonian 
\begin{equation}
\begin{split}
  \hat{H}[n] &= \hat{t}_{SR} + \hat{v}_{ext} + \hat{v}_{es} + \hat{v}_{xc} + \hat{v}_{SOC} \\
                &= \hat{t}_{SR} + \hat{v} + \hat{v}_{SOC} \\
                &= \hat{H}_{SR}[n] + \hat{v}_{SOC},
  \label{eq:SOCHamiltonian}
\end{split}
\end{equation}
where $n$ is the electron density obtained from a scalar-relativistic (SR) calculation, $\hat{t}_{SR}$ is the SR kinetic energy operator, $\hat{v}_{ext}$ is the external potential operator, $\hat{v}_{es}$ is the electrostatic or Hartree potential operator of the electrons, $\hat{v}_{xc}$ is the exchange-correlation potential operator, $\hat{H}_{SR}$ is the SR Hamiltonian operator, and $\hat{v}$ is the effective or Kohn-Sham potential operator.  $\hat{v}_{SOC}$ is the spin-orbit coupling operator
\begin{equation}
  \hat{v}_{SOC}=\frac{i}{4c^{2}}\bm{\hat{\sigma}} \cdot \bm{\hat{p}} \hat{v} \times \bm{\hat{p}},
  \label{eq:SOCDef}
\end{equation}
where atomic units are used.  ``Hatted'' quantities indicate operators.  Spacial vector quantities are indicated by boldfaced characters, and scalar quantities (and individual components of vectors) are unbolded.  Here, $\bm{\hat{p}}$ is the momentum operator and $\bm{\hat{\sigma}}$ is the vector spin operator of Pauli matrices, assumed to be polarized along the z axis,
    \begin{equation}
    \hat{\sigma}_{x} = \left[
		\begin{array}{cc}
		0 & 1 \\
		1 & 0 \\
		\end{array}\right],
    \hat{\sigma}_{y} = \left[
		\begin{array}{cc}
		0 & -i \\
		i & 0 \\
		\end{array}\right],
    \hat{\sigma}_{z} = \left[
		\begin{array}{cc}
		1 &  0 \\
		0 & -1 \\
		\end{array}\right].
    \end{equation}
Equation \ref{eq:SOCHamiltonian} is the textbook expression and is itself already an approximation to the more accurate, fully relativistic Dirac-Kohn-Sham equations (see below).  SOC can be added as part of routine computations in many electronic structure codes, but in fact a variety of different approximations to capture the effects of SOC are commonly employed. Some of the many possible approaches include: (i) solving the Dirac-Kohn-Sham equation directly\cite{Opahle03,HEschrig04,DIRAC16,RZhao16}, (ii) including the SOC term in the zero-order regular approximation (ZORA)\cite{EvanLenthe96,PNichols09}, (iii) the non-self-consistent or (iv) self-consistent second-variational method following a self-consistent scalar-relativistic calculation,\cite{AHMacDonald80,LAPWBook} or (v) the direct inclusion of SOC effects into pseudopotentials\cite{LKleinman80,GBBachelet82,ADalCorso05,PNichols09,MDolg11}.  
In the second-variational method, matrix elements for the full Hamiltonian
  \begin{equation}
  \begin{split}
  H_{m\alpha;m'\alpha'} &= \bra{\psi_{m\alpha}\alpha}[\hat{H}_{SR}[n]+\hat{v}_{SOC}]\ket{\psi_{m'\alpha'}\alpha'} \\
  &= \delta_{mm'} \delta_{\alpha\alpha'}\epsilon_{m\alpha} +  \bra{\psi_{m\alpha}\alpha}\hat{v}_{SOC}\ket{\psi_{m'\alpha'}\alpha'},
  \end{split}
  \label{eq:SecondVariationalMethod}
  \end{equation}
are calculated and diagonalized, where $\alpha$ is a spinor, $\psi_{m\alpha}$ is the scalar-relativistic KS eigenvector for energy index $m$ and spin channel $\alpha$, and $\epsilon_{m\alpha}$ is the SR energy eigenvalue of $\psi_{m\alpha}$.  Second-variational SOC is performed as a post-processed correction on a reduced set of SR eigenvectors with size $2N_{states}$, where $N_{states}$ is the number of SR eigenvectors included.  This is in contrast to the first-variational method, in which the problem is solved on the full set of SR eigenvectors with size $2N_{basis}$, which is equivalent up to unitary transformation to calculation and diagonalization of matrix elements of the full Hamiltonian defined on computational basis functions $\varphi_{n}$,
\begin{equation}
  H_{n\alpha;n'\alpha'} = \bra{\varphi_{n}\alpha}[\hat{H}_{SR}[n]+\hat{v}_{SOC}]\ket{\varphi_{n'}\alpha'}.
  \label{eq:FirstVariationalMethod}
\end{equation}
Since the final diagonalization step of the second-variational method is performed on a system with dimension $2N_{states} \ll 2N_{basis}$, the second-variational method constitutes a notable reduction in problem size compared to the first-variational method. In principle, both the first- and the second-variational approach can be employed non-self-consistently (following a self-consistent SR calculation) or self-consistently by iterating over the eigenvectors obtained from the diagonalization of the Hamiltonians (\ref{eq:SecondVariationalMethod}) or (\ref{eq:FirstVariationalMethod}).

Second-variational spin-orbit coupling (non-selfconsistent or self-consistent) is expected to offer a performance advantage over a full two- or four-component treatment from the outset, as it preserves the fundamental symmetries of scalar relativity for the initial self-consistency cycle of an electronic structure calculation.  Scalar-relativistic self-consistency can be obtained with the assumption of spin collinearity and, depending on the system, real algebra.  The more costly complex linear algebra and doubling of the problem size due to spin non-collinearity, implicit in Eqs. (\ref{eq:SOCDef}), (\ref{eq:SecondVariationalMethod}) and (\ref{eq:FirstVariationalMethod}), are then introduced only after the scalar-relativistic SCF cycle has finished.

Non-self-consistent second-variational spin-orbit coupling has an additional performance advantage, as no additional SCF steps are required and spin-orbit coupling is applied only once per calculation.  In particular, the non-self-consistent approach avoids computationally-expensive operations such as additional density and Fock matrix evaluations, which are required for a self-consistent approach. Towards large system sizes, the $O(N^{3})$ diagonalization would become costly and also occurs only once.  On the other hand, the accuracy of non-selfconsistent SOC is expected to decrease with increasing elemental atomic numbers $Z$, as is documented in the literature.\cite{JKunes01,PCarrier04,RSakuma11,IAguilera15} A detailed quantitative assessment is thus needed to gauge the expected reliability of the approach across the periodic table.

In this paper, we provide a broad assessment of the accuracy of first- or second-variational, non-self-consistent (n.s.c.) SOC for all-electron energy band structure calculations, based on numeric atom-centered orbital (NAO) basis sets. NAO basis sets are widely used in electronic structure theory.\cite{FWAverill73,AZunger77,BDelley82,BDelley90,GteVelde01,KKoepernik99,APHorsfield97,OFSankey89,JMSoler02,TOzaki05} Here, we focus on the NAO basis sets provided with the FHI-aims\cite{Blum09} code, a high-accuracy\cite{lejaeghere2016reproducibility,SRJensen17} implementation of electronic structure theory for molecules and solids, suitable for large-scale simulations\cite{LNemec13,ELPA,SVLevchenko15}.  To benchmark the NAO-based n.s.c. approach to SOC, we consider band structures for 103 crystalline solids, incorporating 66 chemical species (up to Po) and using the semi-local PBE\cite{PerdewBurkeErnzerhof96} functional.  The benchmark set includes 45 elemental materials and 21 alkali halides in addition to a group of 37 compound semiconductors.  Band structures considered in this work are provided online via the NOMAD Repository\cite{NOMAD} and are citable via digital object identifiers (DOIs).  These band structures will aid the community in future methods development involving relativistic effects.

  We compare our results to a self-consistent (s.c.) SOC implementation based on (linearized) augmented plane wave\cite{OAAndersen1975} and local orbital\cite{ESjostedt00,GKHMadsen01} ((L)APW+lo) basis sets, which we abbreviate as ``APW s.c. SOC'', in the all-electron code Wien2k\cite{Blaha01}  Optionally, the accuracy of the (L)APW+lo approach may be improved by including Dirac p$^{1/2}$ local orbitals in the second-variational step.\cite{JKunes01,PCarrier04}  We abbreviate this improved handling of SOC as ``APW+p$^{1/2}$ s.c. SOC''.  

Finally, we also investigate the exchange-correlation functional dependence of second-variational SOC by comparing band structures calculated by the PBE and the  short-range screened hybrid exchange-correlation HSE06 functional\cite{Heyd03,Heyd06,AKrukau06} using the NAO basis set.  SOC-related quantities calculated on semi-local and hybrid-functional levels of theory show close agreement for most materials, but important exceptions exist where qualitative differences in the scalar-relativistic band structures yield quantitative differences for SOC-related quantities.  This highlights the need for qualitative accuracy in the underlying (g)KS scalar-relativistic band structure to achieve quantitative accuracy in spin-orbit-coupled calculations.

\section{Background}

In relativistic Kohn-Sham density-functional theory\cite{AKRajagopal73,AKRajagopal78,AHMacDonald79,EEngel02,TSaue02}, the many-particle wavefunction $\Psi$ of the system is rewritten by ansatz to a single Slater determinant of single-particle 4-component wavefunctions $\psi$ interacting with an effective relativistic potential operator $\hat{v}[n,\bm{j}]$, where the dependence on current density $\bm{j}$ arises from covariance.  Each single-particle wavefunction $\psi$ then satisfies the Dirac-Kohn-Sham equation\cite{KGDyall07,TSaue02},
\begin{equation}
	\left[
		\begin{array}{cc}
		\hat{v} & c\bm{\hat{\sigma}}\cdot\bm{\hat{p}} \\
		c\bm{\hat{\sigma}}\cdot\bm{\hat{p}} & \hat{v}-2c^{2} \\
		\end{array}\right]
		\left[
		\begin{array}{c}
		\psi_{L} \\
		\psi_{S} \\
		\end{array}\right] = 
		\epsilon \left[\begin{array}{c}
		\psi_{L} \\
		\psi_{S} \\
		\end{array}\right]
        \label{eq:DiracKohnSham}	
\end{equation}
where we have neglected the current density dependence of $\hat{v}$ to introduce a scalar $\hat{v}$, and the single-particle bispinor $\psi$ is decomposed into individual spinors $\psi_{L}$ and $\psi_{S}$.  In principle, solution of this equation would suffice to cover (almost) all relevant effects in chemistry and condensed matter science.  However, the 4-component approach necessitates additional computational expense and special care regarding basis sets and other quantities (e.g., energy gradients) and is therefore not commonly pursued in most implementations.

  Three energy branches exist for the Dirac-Kohn-Sham equation:  a positive continuum of unbound energy states with $\epsilon \ge 0$, a negative continuum of unbound energy states for finite systems with $\epsilon \le -2c^{2}$, and a discrete spectrum of bound states lying within the $[0,-2c^{2}]$ gap.  The energy needed to couple the bound spectrum with the negative continuum solutions, $\epsilon \approx 2c^{2}$, would allow for electron-positron pair formation. However, this energy scale well exceeds the energy scales of electronic structure theory\cite{WLiu10} for elements of interest in chemistry and materials science (usually $Z\leq$100).  Accordingly, in the ``no-pair'' approximation, the negative continuum spectrum may be viewed as well-separated from the bound spectrum and neglected, and the bound states may be regarded as electron-like.  

  Although Equation \ref{eq:DiracKohnSham} is a set of four differential equations, in the presence of a scalar potential it has only two degrees of freedom.  There exists a coupling operator $\hat{R}$ such that the ``kinetic balance'' condition
\begin{equation}
	\psi_S =  \hat{R}\psi_{L} \equiv \hat{K}\bm{\hat{\sigma}}\cdot\bm{\hat{p}}\psi_{L}
    \label{eq:KineticBalance}.
\end{equation}
is satisfied for some operator $\hat{K}$, yielding an equation for $\psi_L$ independent of $\psi_S$
\begin{equation}
	[c\bm{\hat{\sigma}}\cdot\bm{\hat{p}}\hat{K}\bm{\hat{\sigma}}\cdot\bm{\hat{p}} + \hat{v}]\psi_L = \epsilon\psi_L.
    \label{eq:DiracPsiL}
\end{equation}
By simply using Equation \ref{eq:DiracKohnSham}, $\hat{K}$ can written down exactly in closed form,
\begin{equation}
	\hat{K} = \frac{1}{2c}(1+\frac{\epsilon-\hat{v}}{2c^{2}})^{-1},
    \label{eq:ExactCoupling}
\end{equation}
for electron-like solutions, with a similar equation holding for positron-like solutions.  However, the explicit dependence on $\epsilon$ complicates the solution, and various further approximation schemes\cite{KGDyall07,TSaue11} exist for $\hat{R}$.

Equation \ref{eq:DiracPsiL} alone does \emph{not} reduce Dirac-Kohn-Sham theory to a two-component formalism, as $\psi_{S}$ is non-zero for non-trivial solutions and contributes to the normalization condition
\begin{equation}
	1 = \braket{\psi|\psi} = \braket{\psi_{L}|\psi_{L}} + \braket{\psi_{S}|\psi_{S}}
    \label{eq:Normalization}
\end{equation}
defined on the full 4-component wavefunction. Elimination of $\psi_{S}$ can be accomplished by exact 2-component formalisms\cite{MReiher09,WLiu10,TSaue11} which fully decouple $\psi_{L}$ and $\psi_{S}$ using numerical construction of $R$ and suitable unitary transformation of Equation \ref{eq:DiracKohnSham}, but this formalism will not be considered in this paper.

  The Dirac identity, valid for operators $\bm{\hat{a}}$ which satisfy $[\bm{\hat{a}},\bm{\hat{\sigma}}] = 0$, asserts that
\begin{equation}
  (\bm{\hat{\sigma}}\cdot\bm{\hat{a}})(\bm{\hat{\sigma}}\cdot\bm{\hat{b}}) = \bm{\hat{a}}\cdot\bm{\hat{b}}\bm{\hat{I}_{2}} + i\bm{\hat{\sigma}}\cdot\bm{\hat{a}}\times\bm{\hat{b}}.
\end{equation}
  Making the identification $\bm{\hat{a}}\equiv\bm{\hat{p}}\hat{K}$ and $\bm{\hat{b}}\equiv\bm{\hat{p}}$, Equation \ref{eq:DiracPsiL} becomes
\begin{equation}
	[c\bm{\hat{p}}\hat{K}\cdot\bm{\hat{p}} + ic\bm{\hat{\sigma}}\cdot\bm{\hat{p}}\hat{K}\times\bm{\hat{p}} + \hat{v}]\psi_L = \epsilon\psi_L
    \label{eq:DiracLargeComp3}
\end{equation}
where $\times$ is the usual cross product between spacial vectors.  Equation \ref{eq:DiracLargeComp3} disentangles the two major relativistic corrections to the non-relativistic Schr\"{o}dinger equation considered in condensed matter physics and quantum chemistry.  The first term, the scalar-relativistic kinetic energy term $c\bm{\hat{p}}\hat{K}\cdot\bm{\hat{p}}$, is independent of spin and modifies the non-relativistic kinetic energy operator $\bm{\hat{p}}^2/2$ by a relativistic renormalization factor $2c\hat{K}$.  The second term couples the spin operator acting on spinors to spacial operators operating on spacial degrees of freedom.  This second term is accordingly called the ``spin-orbit coupling'' term.
 
  For states weakly affected by relativity, e.g. valence and conduction states, the contribution of $\psi_{S}$ to the normalization condition (\ref{eq:Normalization}) is weak and may be neglected in the non-relativistic limit.  $\psi_{L}$ may then be solved and normalized independently of $\psi_{S}$, functioning essentially as the non-relativistic spinor.  From here on, we will rewrite $\psi_{L} \rightarrow \psi$, which we identify as the Schr\"{o}dinger-like 2-component spinor. We will, however, return briefly to the small component $\psi_{S}$ in the context of free-atom orbitals.
  
    Though both the SR kinetic energy term and the SOC term arise from relativistic modification of the kinetic energy operator in the Dirac equation, the SR kinetic energy term is often considered the stronger relativistic effect in electronic structure theory.  Relativistic effects are mostly significant in the nuclear region, where a highly negative $v$ induces large kinetic energy densities via the relation $\hat{t}\psi = (\epsilon - \hat{v})\psi$ even for the lightest atoms.  Accordingly, Schr\"{o}dinger-like atomic $l$=0 orbitals are most strongly affected by scalar-relativistic effects, which diminish in strength for orbitals of increasing $l$, as the electronic density of the orbital spreads further away from the nuclear region of the system.  
    
    In contrast, the spin-orbit coupling term is anti-symmetric in $\bm{\hat{p}}$ due to the presence of the cross product.  Its effect is formally zero on the spherically-symmetric Schr\"{o}dinger atomic $l=0$ orbital.  Higher $l$ orbitals have zero electron density at the nuclei, requiring a compensating increase in the potential energy near the nucleus $v\approx -Z/r$  for SOC to overcome the nodal structure of the Schr\"{o}dinger orbitals and have an appreciable effect.  This gives rise to the well-known dependence of SOC-related effects on the atomic number $Z$ of atoms in a molecule or solid.
    
    The expected difference in strength between these two relativistic corrections suggests an approximation where the two relativistic terms are conceptually decoupled from one another\cite{DDKoelling77} and may be treated by different approximations, 
\begin{equation}
	[c\bm{\hat{p}}\hat{K}_{SR}\cdot\bm{\hat{p}} + ic\bm{\hat{\sigma}}\cdot\bm{\hat{p}}\hat{K}_{SOC}\times\bm{\hat{p}} + \hat{v}]\psi = \epsilon\psi.
    \label{eq:DiracLargeComp4}
\end{equation}
$\hat{K}_{SR}$ refers to the approximation for $\hat{K}$ used in the scalar-relativistic term, and $\hat{K}_{SOC}$ refers to the approximation for $\hat{K}$ used in the SOC term.  
   Using Equation \ref{eq:ExactCoupling}, the lowest-order approximation for $\hat{K}_{SOC}$ is
\begin{equation}
  \hat{K}_{SOC} \approx \frac{1}{2c}.
\end{equation}   
In this approximation, the SOC term in Equation \ref{eq:DiracLargeComp4} is proportional to $\hat{\sigma} \cdot \bm{\hat{p}}\times\bm{\hat{p}} = 0$.  This eliminates the SOC term and yields the scalar-relativistic Schr\"{o}dinger equation,
\begin{equation}
	[c\bm{\hat{p}}\hat{K}_{SR}\cdot\bm{\hat{p}} + \hat{v}]\psi = \epsilon\psi.
    \label{eq:SRSchrodinger}
\end{equation}
The SR Schr\"{o}dinger equation is the workhorse equation of electronic structure theory.  For suitable choice of $\hat{K}_{SR}$, this equation has the same symmetries as the non-relativistic Schr\"{o}dinger equation.  Most of the theoretical and conceptual machinery of the non-relativistic quantum theory can be imported into analysis of the SR Schr\"{o}dinger equation, while still including the dominant relativistic correction for lighter materials to energy levels.
Several different successful approximations for $\hat{K}_{SR}$ exist in the literature. In the present work, two specific variants of $\hat{K}_{SR}$ will be considered (outlined in Sec. \ref{sec:computational} below): (1) the ``atomic zero-order regular approximation'' (atomic ZORA) as used in the FHI-aims code\cite{Blum09} and a combination of a Dirac code and the Koelling-Harmon approximation\cite{DDKoelling77} for valence states as implemented in Wien2k.\cite{Blaha01}

  To reintroduce SOC, Equation \ref{eq:ExactCoupling} is taken to first order in $(\epsilon-\hat{v})/c^2$, 
\begin{equation}
  \hat{K}_{SOC} \approx \frac{1}{2c}-\frac{\epsilon-\hat{v}}{4c^3},
\end{equation}
and inserted in Equation \ref{eq:DiracLargeComp4}, yielding the spin-orbit-coupled Schr\"{o}dinger equation
\begin{equation}
  [c\bm{\hat{p}}\hat{K}_{SR}\cdot\bm{\hat{p}} + \frac{i}{4c^2}\bm{\hat{\sigma}} \cdot \bm{\hat{p}}\hat{v}\times\bm{\hat{p}} + \hat{v}]\psi = \epsilon\psi
  \label{eq:SOCSchrodinger}
\end{equation}

\begin{figure}[H]
    \includegraphics[trim = 0cm 3cm 0cm 3cm, clip, width=0.5\textwidth]{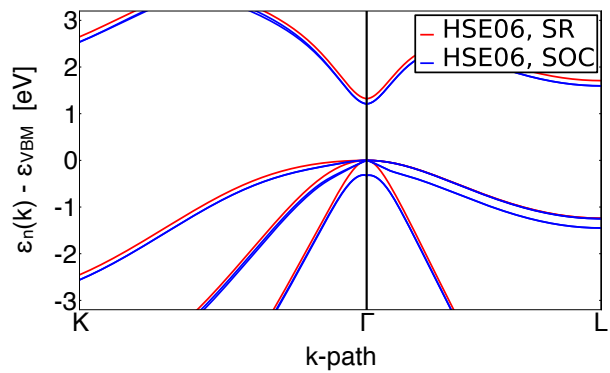}
	\caption{SR and SOC band structures of zincblende GaAs near $\Gamma$ for the HSE06 functional, calculated using FHI-aims with a 16$\times$16$\times$16 k-grid and the ``tight'' NAO basis set.  ``really tight'' integration settings of the FHI-aims code (see Section \ref{sec:aims_desc}) were used.}
	\label{fig:CanonicalBandSplitting}
\end{figure}

  The main effect of SOC on energy levels of materials is the splitting of energy levels that were predicted to be degenerate in SR electronic structure theory.  Figure \ref{fig:CanonicalBandSplitting} illustrates this effect for the split valence band of the (cubic) zincblende compound semiconductor GaAs near the $\Gamma$ point.  This effect, known as spin-orbit splitting, can be understood as a Zeeman-like effect owing to the form of the perturbing SOC operator, and formally as the change in the irreducible representations of the Hamiltonian when transitioning from the single group of non-relativistic/scalar-relativistic quantum theory to the double group of relativistic quantum theory.\cite{KGDyall07,MSDresselhaus08}

\section{Evaluation of the Spin-Orbit Coupling Operator}

\subsubsection{Calculation of Matrix Elements of the Spin-Orbit Coupling Operator}

We next summarize the computational approach to evaluate the spin-orbit-coupled Hamiltonian matrix elements $H_{m\alpha;m'\alpha'}$ in the second-variational method and using a localized basis set. Repeating Eq. (\ref{eq:SecondVariationalMethod}), 
\begin{equation}
  H_{m\alpha;m'\alpha'} = \delta_{mm'} \delta_{\alpha\alpha'}\epsilon_{m\alpha} + \bra{\psi_{m\alpha}\alpha}\hat{v}_{SOC}\ket{\psi_{m'\alpha'}\alpha'} , 
  \label{eq:HSOC1}
\end{equation}
 where $\epsilon_{m\alpha}$ is the SR eigenvalue for the self-consistent SR eigenstate $\psi_{m\alpha}$ with energy index $m$ and spin channel $\alpha$.  For periodic boundary conditions, $H_{m\alpha;m'\alpha'}$, $\psi_{m\alpha}$, and $\epsilon_{m\alpha}$ acquire a $\bm{k}$ dependence by way of the Bloch theorem, and Eq.  (\ref{eq:HSOC1}) must be evaluated for each $\bm{k}$.

  	The SOC operator $\hat{v}_{SOC}$, when explicitly decomposed in terms of its spacial components, has the form
\begin{multline}
	\hat{v}_{SOC} = \frac{-i}{4c^{2}}[(\frac{d}{dy} \hat{v} \frac{d}{dz} - \frac{d}{dz} \hat{v} \frac{d}{dy})\hat{\sigma}_{x} \\
	                                          + (\frac{d}{dz} \hat{v} \frac{d}{dx} - \frac{d}{dx}\hat{v}\frac{d}{dz})\hat{\sigma}_{y} \\
	                                          + (\frac{d}{dx} \hat{v} \frac{d}{dy} - \frac{d}{dy}\hat{v}\frac{d}{dx})\hat{\sigma}_{z}] \\
									\equiv \frac{-i}{4c^{2}}\sum_{x_{i}}\hat{\Pi}_{x_{i}}\hat{\sigma}_{x_{i}}.
\end{multline}
($x_{i}=x,y,z$).  Each term in this decomposition is a tensor product of a spacial operator $\hat{\Pi}_{x_{i}}$, which acts on the spacial wavefunctions $\ket{\psi_{m\alpha}}$, and a spin operator $\hat{\sigma}_{x_{i}}$, which acts on the spinors $\ket{\alpha}$.  The resulting matrix elements in the basis of SR eigenfunctions have the form
\begin{multline}
	\bra{\psi_{m\alpha}\alpha}\hat{v}_{SOC}\ket{\psi_{m'\alpha'}\alpha'} = \\
    \frac{-i}{4c^{2}}\sum_{x_{i}}\bra{\psi_{m\alpha}}\hat{\Pi}_{x_{i}}\ket{\psi_{m'\alpha'}}\bra{\alpha}\hat{\sigma}_{x_{i}}\ket{\alpha'}.
	\label{eq:SOCbasis}
\end{multline}

  Assuming polarization of $\alpha$ in the z-direction ($\ket{\alpha}=\ket{\uparrow},\ket{\downarrow}$,) the matrix elements of $\hat{v}_{SOC}$ assume a block form with respect to the spinor basis via the decomposition
\begin{equation}
\begin{split}
  \hat{v}_{SOC}&=\left[
  \begin{array}{cc}
  \hat{v}_{SOC,\uparrow\uparrow} & \hat{v}_{SOC,\uparrow\downarrow} \\
  \hat{v}_{SOC,\downarrow\uparrow} & \hat{v}_{SOC,\downarrow\downarrow} \\
  \end{array}\right] \\
  &=\frac{-i}{4c^{2}}\left[
  \begin{array}{cc}
  \hat{\Pi}_{z} & \hat{\Pi}_{x} - i\hat{\Pi}_{y} \\
  \hat{\Pi}_{x} + i\hat{\Pi}_{y} & -\hat{\Pi}_{z} \\
  \end{array}\right].
  \label{eq:HSOC2}
\end{split}
\end{equation}
Within each sub-block, one only considers operations on spacial wavefunctions of form $\bra{\psi_{m\alpha}}\hat{\Pi}_{x_{i}}\ket{\psi_{m'\alpha'}}$.

\subsubsection{Calculation of the Spacial Components of SOC Matrix Elements}

\paragraph{Non-Periodic Case}

  The spacial component of a non-periodic eigenvector $\ket{\psi_{m\alpha}}$ defined on a set of localized spacial basis functions $\varphi_{n}$ is representable as a linear combination 
\begin{equation}
	\ket{\psi_{m\alpha}} = \sum_{n}c_{m\alpha n}\ket{\varphi_{n}},
     \label{eq:EigenfunctionBlochStatesNonPeriodic}
\end{equation}

The matrix elements for the $\hat{\Pi}_{x_{i}}$ operators are
\begin{equation}
  \bra{\psi_{m\alpha}}\hat{\Pi}_{x_{i}}\ket{\psi_{m'\alpha'}} \\
      = \sum_{n,n'}c_{m\alpha n}^{*}M_{x_{i};nn'}c_{m'\alpha'n'}
  \label{eq:Pi_1}
\end{equation}
where
\begin{eqnarray}
  M_{x_{i};nn'} & = & \bra{\varphi_{n}}\hat{\Pi}_{x_{i}}\ket{\varphi_{n'}}
  \label{eq:SOC_Matrix_Non_Periodic} \\
  & = & \int \varphi_{n}(\bm{r})\hat{\Pi}_{x_{i}}\varphi_{n'}(\bm{r})d\bm{r}. \nonumber
\end{eqnarray}
$M_{x_{i};nn'}$ depends only on the spacial basis set and needs to be evaluated only once for a given $\hat{v}$. It can be evaluated efficiently using the same real-space, linear-scaling integration scheme as described for FHI-aims' Hamiltonian matrix elements in Ref.\cite{Havu09} 

\paragraph{Periodic Case}
   The spacial component of a periodic eigenvector $\ket{\psi_{m\alpha}(\bm{k})}$ may be written in terms of Bloch basis functions $\ket{\chi_{n}(\bm{k})}$
\begin{equation}
  \ket{\chi_{n}(\bm{k})} = \frac{1}{\sqrt{N_{cell}}}\sum_{\bm{T}}e^{i\bm{k}\cdot\bm{T}}\ket{\varphi_{n,\bm{T}}},
  \label{eq:BlochDef}
\end{equation}
where $\ket{\varphi_{n,\bm{T}}}$ is the periodic image of the n$^{th}$ spacial basis function in the unit cell indexed by $\bm{T}$ and $\lim_{N_{cell}\rightarrow\infty}$ is implicit.  The eigenvectors then acquire a $\bm{k}$ index, 
\begin{equation}
  \ket{\psi_{m\alpha}(\bm{k})} = \sum_{n}c_{m\alpha n}(\bm{k})\ket{\chi_{n}(\bm{k})} .
  \label{eq:EigenfunctionBlochStates}
\end{equation}
    
  Using Equation \ref{eq:BlochDef} and \ref{eq:EigenfunctionBlochStates}, the matrix elements for the $\hat{\Pi}_{x_{i}}$ operators are
\begin{multline}
  \bra{\psi_{m\alpha}(\bm{k})}\hat{\Pi}_{x_{i}}\ket{\psi_{m'\alpha'}(\bm{k})} \\
      = \frac{1}{N_{cell}}\sum_{n,n'}c_{m\alpha n}^{*}(\bm{k})c_{m'\alpha'n'}(\bm{k}) \times \\
      \sum_{\bm{TT'}}e^{i\bm{k}\cdot(\bm{T'}-\bm{T})}\bra{\varphi_{n,\bm{T}}}\hat{\Pi}_{x_{i}}\ket{\varphi_{n',\bm{T'}}}
\label{eq:PeriodicPi}
\end{multline}    
  By translational symmetry,
\begin{equation}
\begin{split}
	\bra{\varphi_{n,\bm{T}}}\hat{\Pi}_{x_{i}}\ket{\varphi_{n',\bm{T'}}} &\equiv \bra{\varphi_{n,\bm{T}}}\hat{\Pi}_{x_{i}}\ket{\varphi_{n',\bm{\tau}+\bm{T}}} \\
      &= \bra{\varphi_{n,\bm{0}}}\hat{\Pi}_{x_{i}}\ket{\varphi_{n',\bm{\tau}}},
\end{split}
\end{equation}
    that is, we may rewrite these matrix elements without loss of generality in terms of a ``central unit cell'' (with index $\bm{T}=0$) and index $\bm{\tau}=\bm{T'}-\bm{T}$ relative to the central unit cell.  Equation \ref{eq:PeriodicPi} then becomes
\begin{multline}
	\bra{\psi_{m\alpha}(\bm{k})}\hat{\Pi}_{x_{i}}\ket{\psi_{m'\alpha'}(\bm{k})} \\
      = \frac{1}{N_{cell}}\sum_{n,n'}c_{m\alpha n}^{*}(\bm{k})c_{m'\alpha'n'}(\bm{k})\times \\
        \sum_{\bm{T\tau}}e^{-i\bm{k}\cdot\bm{\tau}}\bra{\varphi_{n,\bm{0}}}\hat{\Pi}_{x_{i}}\ket{\varphi_{n',\bm{\tau}}} \\
      = \sum_{n,n'}c_{m\alpha n}^{*}(\bm{k})c_{m'\alpha'n'}(\bm{k})\sum_{\bm{\tau}}e^{-i\bm{k}\cdot\bm{\tau}}\bra{\varphi_{n,\bm{0}}}\hat{\Pi}_{x_{i}}\ket{\varphi_{n',\bm{\tau}}} \\
      =  \sum_{n,n'}c_{m\alpha n}^{*}(\bm{k})[\sum_{\tau}e^{-i\bm{k}\cdot\bm{\tau}}\bra{\varphi_{n,\bm{0}}}\hat{\Pi}_{x_{i}}\ket{\varphi_{n',\bm{\tau}}}]c_{m'\alpha'n'}(\bm{k})
  \label{eq:Periodic_1}
\end{multline}
   
   Equation \ref{eq:Periodic_1} may be cast in a form analogous to the non-periodic case,
   \begin{multline}
	\bra{\psi_{m\alpha}(\bm{k})}\hat{\Pi}_{x_{i}}\ket{\psi_{m'\alpha'}(\bm{k})} \\ =
      \sum_{n,n'}c_{m\alpha n}^{*}(\bm{k})W_{x_{i};nn'}(\bm{k})c_{m'\alpha'n'}(\bm{k}).
    \label{eq:Periodic_2}
\end{multline}
   Here, 
\begin{equation}
W_{x_{i};nn'}(\bm{k}) = \sum_{\tau}e^{-i\bm{k}\cdot\bm{\tau}}M_{x_{i};nn'\tau}
  \label{eq:Work_Matrix}
\end{equation}
incorporates all interactions between a given localized basis function $\varphi_{n}$ in the central cell and all periodic images of $\varphi_{n'}$ with overlapping support, and
   \begin{equation}
  M_{x_{i};nn'\tau}=\bra{\varphi_{n,\bm{0}}}\hat{\Pi}_{x_{i}}\ket{\varphi_{n',\bm{\tau}}},
  \label{eq:SOC_Matrix_Periodic}
\end{equation}
is the periodic analogue of $M_{x_{i};nn'}$.  

  $M_{x_{i};nn'\tau}$ is independent of $\bm{k}$ and depends only on the spacial basis, so it is evaluated once for a given $\hat{v}$.  The matrix $W_{x_{i};nn'}(\bm{k})$ is evaluated once for each k-point $\bm{k}$.  Once $W_{x_{i};nn'}(\bm{k})$, is computed, solution of Equation \ref{eq:Periodic_2} requires two matrix multiplications for each $\bm{k}$.
    
\paragraph{Avoiding explicit derivatives of the potential.} Consider $M_{z;nn'}$ in the non-periodic case, where it has the form
\begin{equation}
	M_{z;nn'} = \bra{\varphi_{n}}\frac{d}{dx} \hat{v} \frac{d}{dy}\ket{\varphi_{n'}} - \bra{\varphi_{n}}\frac{d}{dy} \hat{v} \frac{d}{dx}\ket{\varphi_{n'}}.
\end{equation}
This matrix is anti-symmetric, with the hermiticity of the SOC operator $\hat{v}_{SOC}$ preserved by the strictly imaginary coefficient of the overall operator.  For localized basis functions, this matrix element can be rewritten by integration by parts to read 
\begin{equation}
	M_{z;nn'} = - \bra{\frac{d\varphi_{n}}{dx}} \hat{v} \ket{\frac{d\varphi_{n'}}{dy}} + \bra{\frac{d\varphi_{n}}{dy}} \hat{v} \ket{\frac{d\varphi_{n'}}{dx}}.
   \label{eq:DoubleGradients}
\end{equation}
due to anti-hermiticity of the gradient operator, with similar forms for the matrix elements of $\hat{\Pi}_{x}$ and $\hat{\Pi}_{y}$.  An analogous result holds for the periodic case.

By using Equation \ref{eq:DoubleGradients}, gradients of the potential $\nabla \hat{v}$ do not need to be calculated.  The gradients of the wavefunctions of basis elements $\nabla \varphi_{n,\tau}$ needed can be readily computed in any electronic structure code based on localized basis functions.

\subsubsection{Steps for Second-Variational SOC in a Localized Basis Set}
  To summarize, we briefly outline the steps for incorporating SOC in a second-variational approach. \\
  
\paragraph*{Non-Periodic Case}
\begin{enumerate}
    \item Perform a self-consistent scalar-relativistic calculation to obtain SR eigenfunctions $\psi_{m\alpha}$ and eigenvalues $\epsilon_{m\alpha}$.
    \item Select a subset of SR eigenvectors $\psi_{m\alpha}$ for inclusion in the second-variational method.
    \item Calculate $M_{x_{i};nn'}$, the matrix elements of the operators $\hat{\Pi}_{x_{i}}$ between localized basis functions $\varphi_{n}$, via Equation \ref{eq:SOC_Matrix_Non_Periodic}.
    \item Use Equation \ref{eq:Pi_1} to calculate the matrix elements of $\hat{\Pi}_{x_{i}}$ between the scalar-relativistic eigenvectors chosen in Step 2.
    \item Use Equation \ref{eq:SOCbasis} and Equation \ref{eq:HSOC1} to construct the Hamiltonian matrix elements $H_{m\alpha;m'\alpha'}$ of the system.
    \item Diagonalize $H_{m\alpha;m'\alpha'}$ to obtain the resulting SOC eigenvalues and (if needed) eigenvectors of the system.
\end{enumerate}

\paragraph*{Periodic Case}
\begin{enumerate}
    \item Perform a self-consistent scalar-relativistic calculation to obtain SR eigenfunctions $\psi_{m\alpha}(\bm{k})$ and eigenvalues $\epsilon_{m\alpha}(\bm{k})$.
    \item Select a subset of SR eigenvectors $\psi_{m\alpha}(\bm{k})$ to include in the second-variational method.
    \item Calculate $M_{x_{i};nn'\tau}$, the matrix elements of the operators $\hat{\Pi}_{x_{i}}$ between localized basis functions $\varphi_{n,\bm{T}}$, via Equation \ref{eq:SOC_Matrix_Periodic}.
    \item For each k-point $\bm{k}$, use Equation \ref{eq:Work_Matrix} to calculate the matrix $W_{x_{i};nn'}(\bm{k})$.
    \item For each k-point $\bm{k}$, use Equation \ref{eq:Periodic_2} to calculate the matrix elements of $\hat{\Pi}_{x_{i}}$ between the scalar-relativistic eigenvectors chosen in Step 2.
    \item For each k-point $\bm{k}$, use Equation \ref{eq:SOCbasis} and Equation \ref{eq:HSOC1} to construct the Hamiltonian matrix elements $H_{m\alpha;m'\alpha'}(\bm{k})$ of the system at $\bm{k}$.
    \item For each k-point $\bm{k}$, diagonalize $H_{m\alpha;m'\alpha'}(\bm{k})$ to obtain the resulting SOC eigenvalues and (if needed) eigenvectors of the system at $\bm{k}$.
\end{enumerate}

In practice, our NAO-based implementation in FHI-aims includes one further approximation, which is to omit $\hat{v}_{xc}$ from the effective potential operator $\hat{v}$ appearing in $\hat{v}_{SOC}$. The effect of this omission is small since relativistic effects only contribute significantly to the Hamiltonian deep in the nuclear region. This simplification yields a form for the SOC operator
\begin{equation}
	\hat{v}_{SOC}\approx \frac{i}{4c^{2}}\bm{\hat{p}} \hat{\tilde{v}} \times \bm{\hat{p}} \cdot \bm{\hat{\sigma}},
\end{equation}
where
\begin{equation}
	\hat{\tilde{v}} \equiv \hat{v} - \hat{v}_{xc} = \hat{v}_{ext} + \hat{v}_{es}.
\end{equation}
This approximation is assessed for the Perdew-Wang parameterization of the local-density approximation functional (PW-LDA)\cite{JPPerdew92} in Table \ref{tab:xc_dependence}, where spin-orbit splittings for select materials are calculated without and with the inclusion of $\hat{v}_{xc}$ in the effective potential operator $\hat{v}$ used for $\hat{v}_{SOC}$.  The spin-orbit splitting at VBM for a given material is presented, with the exception of the Pb-containing compounds where the spin-orbit splitting in the first conduction band at $\Gamma$ is presented.  Calculations were performed using FHI-aims at ``Benchmark Settings'', described in Section \ref{sec:aims_desc}.  

We find that inclusion of $\hat{v}_{xc}$ changes spin-orbit splittings by 1\% for these materials.  While the omission of $\hat{v_{xc}}$ does not affect the previous derivation conceptually, in practice it simplifies the implementation (higher derivatives of gradient-based exchange-correlation functionals that would be needed for the potential expression are avoided) and also allows one to use a local potential operator in calculations involving hybrid functionals.

\begin{table}
\centering
\begin{tabular}{|c|c|c|c|}
\hline
Material & Prototype &    VBM Splitting      &   VBM Splitting  \\ 
         &           & No $v_{XC}$ in $v_{soc}$ & $v_{XC}$ in $v_{soc}$ \\    
\hline
\hline
ZnS       & ZB & 0.064 & 0.065 \\
\hline
GaP       & ZB & 0.089 & 0.090 \\
AlAs      & ZB & 0.312 & 0.315 \\
GaAs      & ZB & 0.350 & 0.353 \\
ZnSe      & ZB & 0.403 & 0.406 \\
KBr       & RS & 0.517 & 0.522 \\
RbCl      & RS & 0.137 & 0.140 \\
CdS       & ZB & 0.047 & 0.048 \\
CdSe      & ZB & 0.382 & 0.385 \\
\hline
InP       & ZB & 0.108 & 0.109 \\
AlSb      & ZB & 0.687 & 0.691 \\
NaI       & RS & 1.107 & 1.113 \\
CsF       & RS & 0.157 & 0.160 \\
HgS       & ZB & 0.692 & 0.692 \\
\hline
\hline
Material & Prototype &    CB Splitting     &   CB Splitting \\ 
         &           & No $v_{XC}$ in $v_{soc}$ & $v_{XC}$ in $v_{soc}$ \\    
\hline
PbS       & RS & 2.793 & 2.804 \\
PbSe      & RS & 2.619 & 2.629 \\
PbTe      & RS & 2.350 & 2.359 \\
\hline
\end{tabular}
\caption{A comparison of select spin-orbit splittings using the PW-LDA functional, with and without the XC contribution to the effective potential operator used in $\hat{v}_{SOC}$.  Values are presented in units of eV. ``RS'' and ``ZB'' denote rocksalt and (cubic) zincblende prototypes, respectively.}
\label{tab:xc_dependence}
\end{table}

\section{Computational details}
\label{sec:computational}
\subsection{Materials and k-Paths for Band Structures}
	We use the band structures of 103 materials for benchmarking the implementation of spin-orbit coupling.  A brief overview of these materials is presented in Table \ref{tab:Materials}.  Structural prototypes and the lattice parameters used may be found in Tables 3-5 of the Supplemental Material (SM). 
\begin{table}
\centering
\begin{tabular}{|c|c|c|}
\hline
Family & \# Materials & Materials \\    
\hline
Elementals     &  45   & \underline{Be}, \underline{C} [GRA], Ne, \underline{Mg}, \\
               &       & Al, Si, \underline{Ca}, \underline{Sc}, \underline{Ti}, V, \\
               &       & Cr, Mn, Fe, Co, Ni, \\
               &       & Cu, \underline{Zn}, Ge, Sr, Y, \\
               &       & Zr, Nb, Mo, Tc, Ru, \\
               &       & Rh, Pd, Ag, \underline{Cd}, Sn, \\
               &       & Xe, Ba, Lu, Hf, Ta, \\
               &       & W, Re, Os, Ir, Pt, \\
               &       & Au, Tl, Pb, Bi, Po \\
\hline
Compound       &  37   & \underline{C} [DIA], \underline{MgO}, \underline{AlN} [WUR],  \\
Semiconductors &       & \underline{AlN} [ZB], \underline{SiC}, \underline{BP}, AlP,  \\
               &       & MgS, ZnO, ZnS [WUR], \\ 
               &       & ZnS [ZB], GaN [WUR], \\
               &       & \underline{GaN} [ZB], GaP, AlAs, \\
               &       & BAs, GaAs, MgSe, ZnSe, \\
               &       & CdS [WUR], CdS [ZB], \\ 
               &       & CdSe [WUR], CdSe [ZB], \\
               &       & InN, InP, InAs, AlSb, \\
               &       & GaSb, InSb, ZnTe, CdTe, \\
               &       & HgS, HgSe, HgTe, \\
               &       & PbS, PbSe, PbTe \\
\hline
Alkali         &  21   & LiF, NaF, LiCl, NaCl, \\
Halides        &       & KF, KCl, LiBr, NaBr, \\
               &       & KBr, RbF, RbCl, RbBr, \\
               &       & LiI, NaI, KI, RbI, \\
               &       & CsF, CsCl [CSCL], \\
               &       & CsCl [RS], CsBr, CsI \\ 
\hline
\end{tabular}
\caption{Materials used in this study, grouped by type. For materials with identical chemical composition but differing prototypes, the prototype has been indicated in brackets (see Section 3 of the SM).  Underlined materials were used for the ``band delta'' benchmark (Section \ref{sec:band_delta}) but not in the comparison of spin-orbit splittings.}
\label{tab:Materials}
\end{table}

For most materials, we use experimental lattice parameters taken from Pearson's Handbook\cite{Pearson1985}.  Notable exceptions are noble gas solid phases, for which lattice parameters are taken from Villars and Daams\cite{PVillars93}, and alkali halides, for which lattice parameters are taken from Wyckoff\cite{Wyckoff1963}.  
    
   We use the k-paths proposed by Setyawan and Curtarolo\cite{Setyawan10} for all band structures presented, with 21 even-spaced k-points per k-path segment.  The energy zero in band structures is set to the valence band maximum for insulators and to the Fermi level for metals.  All reported energies are in units of eV.  All scalar-relativistic calculations reported in the main text of this paper were performed without spin polarization.  A comparison of the calculated band structures for FCC Ni on the spin-polarized scalar-relativistic and subsequent spin-orbit-coupled levels of theory is provided in Figure S1 in the SM.  Excellent agreement is observed between FHI-aims and WIEN2k on both levels of theory, consistent with the trends observed in the main text.  We briefly summarize code-dependent settings in the following two subsections;  details may be found in Section 2 of the SM.

\subsection{FHI-aims benchmark calculations}
  \label{sec:aims_desc}
  All NAO calculations are performed using FHI-aims\cite{Blum09,Havu09,XRen12,SVLevchenko15,Ihrig2015}, a full-potential all-electron DFT code.  In FHI-aims, basis functions have the form
\begin{equation}
	\varphi_{n}(\bm{r}) = \frac{u_{n}(r)}{r}Y_{lm}(\Omega),
\end{equation}
  where $r$ is the distance from the nucleus of the atom associated with the basis function, $\Omega$ is the solid angle with respect to the associated atom, $u_{n}(r)$ are numerically tabulated functions, and $Y_{lm}(\Omega)$ are real-valued spherical harmonics, with $l$ and $m$ implicitly depending on the index $n$.  $u_{n}(r)$ may be constructed to be exactly zero for $r \ge r_{cut}$, introducing spacial locality and linear scaling (with respect to system size) of computational effort for integrals.\cite{Blum09}
  
  FHI-aims includes preconstructed NAO basis sets for elements $Z$=1-102. These basis sets are organized in so-called ``tiers'' or levels of increasing accuracy (see Table 1 in Ref. \cite{Blum09} and Section 3 in Ref.\cite{SRJensen17_Supp} for several examples) and are constructed for DFT-based total energy calculations. In Ref. \cite{lejaeghere2016reproducibility}, the accuracy of these basis sets was shown to be on par with the best available benchmark codes for calculated scalar-relativistic equations of state for 71 elemental solids. All basis sets include the occupied Kohn-Sham core and valence orbitals for spherical free atoms, known as the ``minimal basis'' in the literature.  Additional basis functions can be added individually or in groups (tiers). The numerical definition of these basis functions is element-specific; for instance, a C atom has three radial functions of $s$, $p$, $d$ character in the first tier, five radial functions of $s$, $p$, $d$, $f$, $g$ character in the second tier, and so on.\cite{Blum09} Additionally, defaults for basic numerical settings (integration grids, Hartree potential, and basis functions) are provided at three levels: ``light'', ``tight'', and ``really\_tight.''  We use ``really\_tight'' integration settings throughout this paper.

For the scalar-relativistic kinetic energy operator of Eq. (\ref{eq:DiracLargeComp4}) in FHI-aims, we use the atomic zero-order regular approximation (``atomic ZORA'') as defined in Eq. (55) of Ref. \cite{Blum09}, and as benchmarked in Ref. \cite{lejaeghere2016reproducibility}  In atomic ZORA, the kinetic energy operator operating on an atom-centered orbital $\varphi_{n}(\bm{r})$ centered about atom $j$ is expressed as
\begin{equation}
	\hat{t}_{at.ZORA}~\varphi_{n}(\bm{r}) = \bm{\hat{p}}\cdot\frac{c^{2}}{2c^{2}-v^{free}_{at(j)}(\bm{r})}\bm{\hat{p}}~\varphi_{n}(\bm{r}),
\end{equation}
where $v^{free}_{at(j)}(\bm{r})$ is the on-site free-atom potential near the nucleus of atom $j$. Since the form of $\hat{t}_{at.ZORA}$ depends explicitly on the atom index of the basis function $\varphi_{n}$ that it acts upon, and since the operator can either act to the left or to the right in a matrix element, the atomic ZORA matrix elements are symmetrized to restore hermiticity.
  
    We use two classes of settings in this paper, described in detail in the Section 2.1 of the SM.  For comparison between NAO and (L)APW+lo data, ``Benchmark Settings'' are used, in which Monkhorst-Pack\cite{MonkhorstPack76} k-grids with k-point densities similar to the $\Delta$ project\cite{KLejaeghere14,lejaeghere2016reproducibility} are used.  Tier 2 basis sets are used in FHI-aims for all Benchmark Settings calculations.  Usually, all possible eigenstates consistent with the basis set are calculated and included in second-variational spin-orbit coupling (exceptions may be found in Section 2.1 of the SM).  Band structures calculated using FHI-aims and Benchmark Settings may be found on the NOMAD Repository via Ref.\cite{Huhn2017BandStructureBenchmarkFHI-aims_BenchmarkSettings}
    
  For comparisons between PBE and HSE06 calculations, ``Tight Production Settings'' are used, in which FHI-aims' tight basis sets are used and $\Gamma$-centered 16x16x16 k-space integration grids are generally employed.  Materials using coarser k-space grids are given in detail in Table 1 of the SM.  The HSE06 functional in this work is defined by a fixed screening parameter ($\omega=0.2$~\AA$^{-1}$) and a fixed exchange mixing parameter ($\alpha$=0.25).  Band structures calculated using FHI-aims and Tight Production Settings may be found on the NOMAD Repository via Ref.\cite{Huhn2017BandStructureBenchmarkFHI-aims_TightProductionSettings}
  
  For materials with maximum atomic numbers smaller than $Z\leq 79$ (Au and lighter), we use FHI-aims' default value for the number of calculated eigenstates in Tight Production Setting, which is given approximately by the empirical formula 
\begin{equation}
  n_{states} \approx n_{electrons} + \sum_{atoms}(2+(1+l_{max,atom})^{2})
\end{equation}  
where $n_{electrons}$ is the number of electrons in the calculation and $l_{max,atom}$ is the maximum (occupied) angular momentum channel for the indicated atom.  In practice, the default value gives roughly 35\%-75\% of all possible eigenstates consistent with the tight basis set and is sufficient to converge second-variational spin-orbit coupling in valence and lowly-lying conduction states.  For materials containing species with $Z\geq 80$ (Hg and heavier), we increase the number of calculated empty states per atom to 50.  For reference, the default \emph{total} number of empty states calculated for HgS is 24.  A study of the effect of basis set size and number of empty states included in second-variational SOC on calculated spin-orbit splittings may be found in Appendix B of the SM.

\subsection{Wien2k benchmark calculations}
  All (L)APW+lo-based calculations are performed using the Wien2k code.\cite{LAPWBook,Blaha01}  In the (L)APW approach, each atom has a ``muffin-tin radius'' $R_{MT}$, defining a sphere that partitions space into two types of regions: core regions S [$r \le R_{MT}$] near each nucleus, where each basis function is represented by a linear combination of atom-centered functions, and an interstitial/valence region I in the rest of space, where each basis function has the form of a single plane wave with wave vector $\bm{k_{\bm{G}}}$, i.e., proportional to  $e^{i\bm{k_{\bm{G}}}\cdot\bm{r}}$.

  In (L)APW+lo, the basis set size is determined by $\frac{1}{2}K_{max}^{2}$, the highest-energy plane wave included in the basis set.  $K_{max}$ is often specified as a product with the smallest muffin-tin radius $R_{MT}$ in the calculation, and this quantity $R_{MT}K_{max}$ serves as an indicator of how many (L)APW+lo basis functions are used in a given run and thus the degree of basis set convergence achieved.  
  
  We present only PBE results using Benchmark Settings for Wien2k.  We generally use the converged ``WIEN2k/acc'' settings from the $\Delta$-project,\cite{KLejaeghere14,lejaeghere2016reproducibility} which were constructed for calculation of total energies of low-temperature elemental structures.  The settings used may be found in Section 2.2 of the SM.   Band structures calculated using WIEN2k and Benchmark Settings may be found on the NOMAD Repository via Ref.\cite{Huhn2017BandStructureBenchmarkWIEN2k_BenchmarkSettings}
  
  The relativistic approximations used in Wien2k vary based on the character of the states.  For core states, a spin-compensated Dirac solver is used.  For semi-core and valence/conduction states in the atomic-sphere region, scalar-relativistic effects are included using the Koelling-Harmon approximation\cite{DDKoelling77} and SOC effects can be included using the self-consistent second-variational method adapted for the (L)APW+lo method.  No relativistic corrections are applied in the interstitial region.  We only consider valence/conduction states in this paper, and accordingly we will refer to the (L)APW+lo basis set as scalar-relativistic.  A cutoff energy of 10 Ry was used, below which all states were included in the second-variational SOC calculation. As mentioned above, Wien2k optionally allows one to include Dirac p$^{1/2}$ local orbitals in the second-variational SOC step, denoted as ``APW+p$^{1/2}$ s.c. SOC'' below. 
  
\section{Results}

\subsection{Comparison of Different Approaches to Relativity and SOC in Spherical Free Atoms}

Since SOC effects arise predominantly in the ``deep'' potential regions near a nucleus, it is instructive to first recall the effects of SOC and relativity on the atomic radial functions and energy levels of moderate to heavy atoms. The spherically symmetric, closed-shell Hg ($Z$=80) atom is chosen to allow for a comparison to exact atomic Dirac radial functions. The reference solutions are orbitals and energies obtained using the fully relativistic Dirac-Kohn-Sham solver dftatom\cite{OCertik13} for spherical free atoms. 
For the exchange-correlation treatment in the atomic calculations, we use the Perdew-Wang parameterization of the local-density approximation functional (PW-LDA).\cite{JPPerdew92}  

\begin{figure*}
  \centering
  \includegraphics[trim={0cm 0.7cm 0cm 1cm},clip,width=\textwidth]{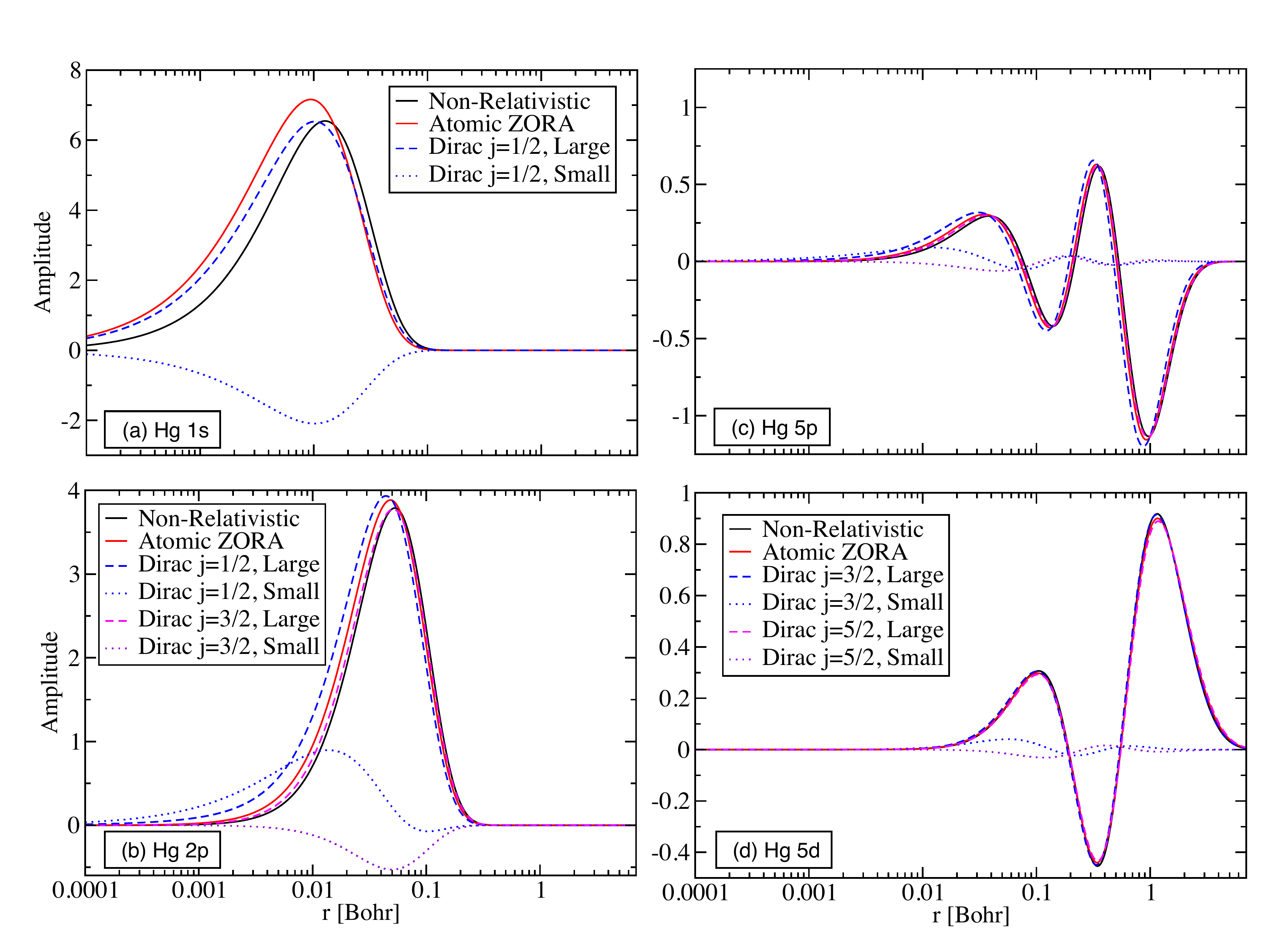}
  \caption{Orbital shapes for the (a) 1s, (b) 2p, (c) 5p, and (d) 5d orbital(s) of a free Hg atom calculated with the PW-LDA functional for various relativistic approximations.  FHI-aims (tier 2) was used for non-relativistic and scalar-relativistic (atomic ZORA) orbitals, and dftatom was used for Dirac orbitals.}
\label{fig:Hg_Orbitals}
\end{figure*}

In Figure \ref{fig:Hg_Orbitals}, the fully relativistic radial functions  of several core (1$s$, 2$p$), semicore (5$p$) and valence (5$d$) Hg orbitals are contrasted with the radial functions from a non-relativistic treatment and from the atomic ZORA scalar-relativistic treatment as implemented in FHI-aims. The scale on the $x$ axes is logarithmic. It is worth noting that, in the area very close to the nucleus, large total energy contributions can originate already from small radial function changes due to the very deep $Z/r$ nuclear Coulomb potential in this range. For the 1$s$ orbital (which cannot show any spin-orbit splitting), the maximum of the non-relativistic radial function occurs at a noticeably larger value than that of the large component radial function. The maximum of the radial function of ZORA lies even further inward. In line with the literature,\cite{EvanLenthe93,EvanLenthe94} the 1$s$ orbital energy of ZORA is therefore significantly lower than the proper Dirac solution, while the non-relativistic 1$s$ orbital energy is higher. However, since the contribution from the core orbital is almost entirely dominated by the nucleus, this particular effect cancels practically exactly in any total-energy difference (as evidenced in  Ref. \cite{lejaeghere2016reproducibility}) or in the SR orbital energy associated with semicore and valence levels. The small component (also shown in Fig. \ref{fig:Hg_Orbitals}a) is not at all negligible, but also cancels in energy differences. It is noteworthy that the overall combined density of the small and large 1$s$ components is closer to the atomic ZORA 1$s$ radial function than the large component on its own.

The 2$p$ core orbital (Fig. \ref{fig:Hg_Orbitals}b) is the radial function with the largest overall SOC effect, since its relativistic version is split into a $j$=1/2 component and a $j$=3/2 component; as is well known,\cite{LAPWBook,JKunes01,PCarrier04} the $j$=1/2 component differs fundamentally from the scalar-relativistic 2$p$ function in that it has no angular momentum barrier and thus a finite probability density to find a 2$p$ electron at the nucleus. For the Hg atom, the split is clearly reflected in the two different components. The non-relativistic 2$p$ function is close to the Dirac $j$=3/2 function but interestingly, the SR ZORA 2$p$ radial function near its maximum resides inbetween the two Dirac 2$p$ components.

The same trend persists into the 5$p$ semicore functions (Fig. \ref{fig:Hg_Orbitals}c), i.e., the ZORA 5$p$ radial function resides between the $j$=3/2 and $j$=1/2 Dirac components. The Dirac 5$p$ functions inherit the visible difference between the radial function components that is already apparent in the 2$p$ functions. This means that a n.s.c. second-variational treatment of SOC following a SR treatment must face limits for the 5$p$ functions, since the second-variational treatment will only approximate their energy difference but will not restore the difference of the underlying $j$=1/2 and $j$=3/2 radial functions themselves. In contrast, the difference between the 5$d$ valence radial functions of Hg in Fig. \ref{fig:Hg_Orbitals}d is much smaller. 

\begin{figure}
  \includegraphics[trim={9.5cm 0cm 9cm 0cm},clip,width=0.5\textwidth]{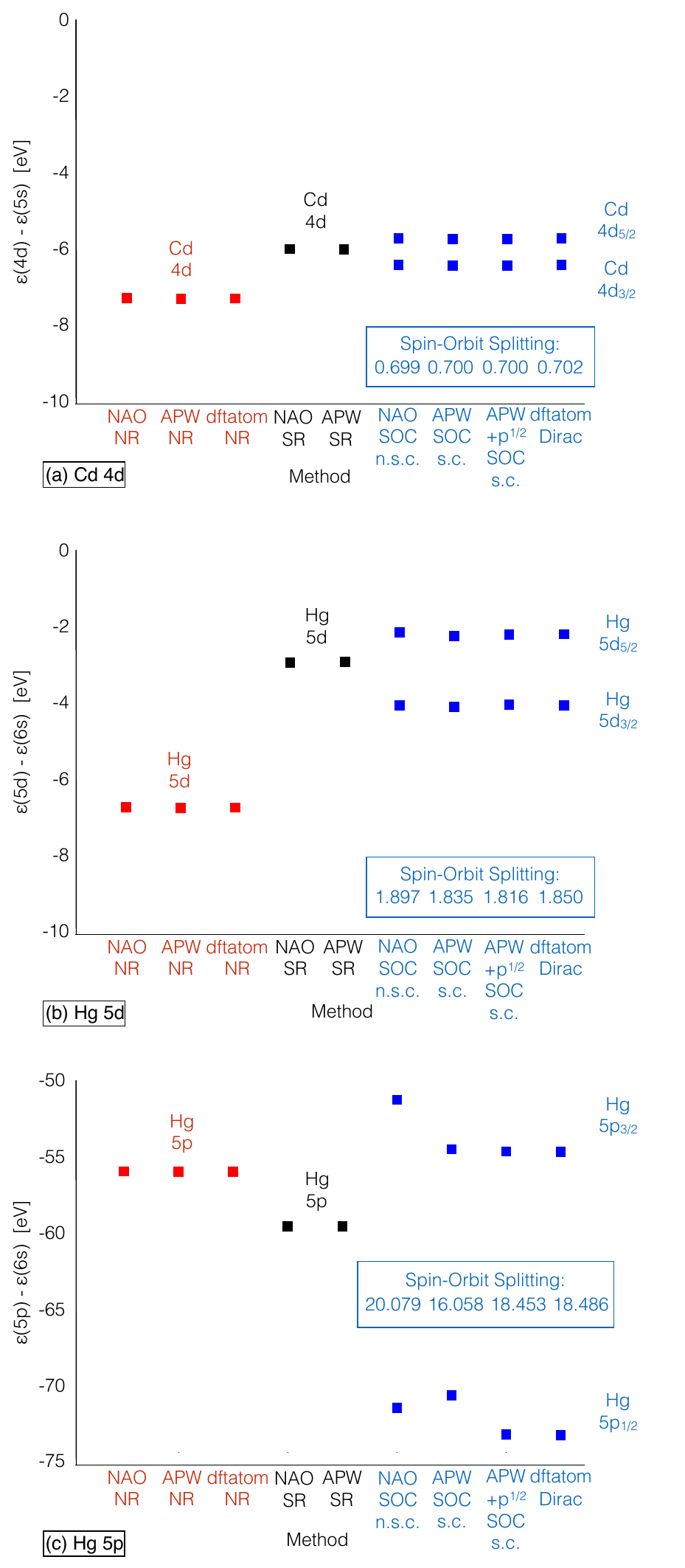}
  \caption{Energy levels of select orbitals of free Cd and Hg atoms calculated with the PW-LDA functional\cite{JPPerdew92} for various relativistic approximations.  Red denotes non-relativistic calculations, black denotes scalar-relativistic calculations, and blue denotes SOC/Dirac-based methods.  Spin-orbit splittings are listed in the inset table, and the energy levels are given in Appendix A of the SM.}
  \label{fig:EnergyLevels}
\end{figure}

These trends manifest themselves quantitatively in Fig. \ref{fig:EnergyLevels}, which compares the actual energy levels of the Cd 4$d$ and the Hg 5$d$ valence orbitals, as well as the Hg 5$p$ semicore orbitals. This figure incorporates energy levels obtained using the Wien2k code, the FHI-aims code, and the dftatom code as a reference. To allow for a comparison of atomic energy levels in periodic and non-periodic calculations, the highest occupied energy levels (5s orbital for Cd, 6s orbital for Hg) are chosen as the energy zero. For Wien2k, the second-variational s.c. (L)APW+lo levels of theory without and with additional p$^{1/2}$ local orbitals are shown. In Wien2k, the free atoms were placed in cubic unit cell with edges $d$=10 \AA, and Benchmark Settings (with $R_{MT}$ increased to 2.26 for Hg) were used.  In FHI-aims, non-periodic calculations with Benchmark Settings were performed. 

For each orbital, Figs. \ref{fig:EnergyLevels}(a-c) include non-relativistic orbital energies for FHI-aims, Wien2k, and dftatom. Since this level of theory is formally identical in all three codes, the differences due to the different numerical implementations are small (within several tens of meV), as expected. The SR treatments in Wien2k and FHI-aims are different regarding the core orbitals, but for the atomic semicore and valence orbitals investigated here, the differences are again small.  

For the spin-orbit split levels, the Dirac solution given by the dftatom code is the appropriate reference. For the valence orbitals, 4$d$ for Cd and 5$d$ for Hg, all three alternative approaches (NAO n.s.c. SOC, APW s.c. SOC, and APW+p$^{1/2}$ s.c. SOC) yield practically identical results at the same scale (differences of a few tens of meV at most) as the NR and SR treatments in all three codes. This is consistent with the observation that the SOC-split $d$ valence radial functions are practically identical in these cases.  In contrast, the Hg 5$p$ semicore orbital is a semicore state, more closely localized near the nucleus than a valence orbital. The 5$p$ $j$=1/2 and $j$=3/2 radial functions are appreciably different, so the three different SOC treatments show noticeable differences as well. 

\subsection{Comparison of Different Approaches to SOC in GaAs and Other Compound Semiconductors}

\begin{figure*}
  \includegraphics[width=\textwidth]{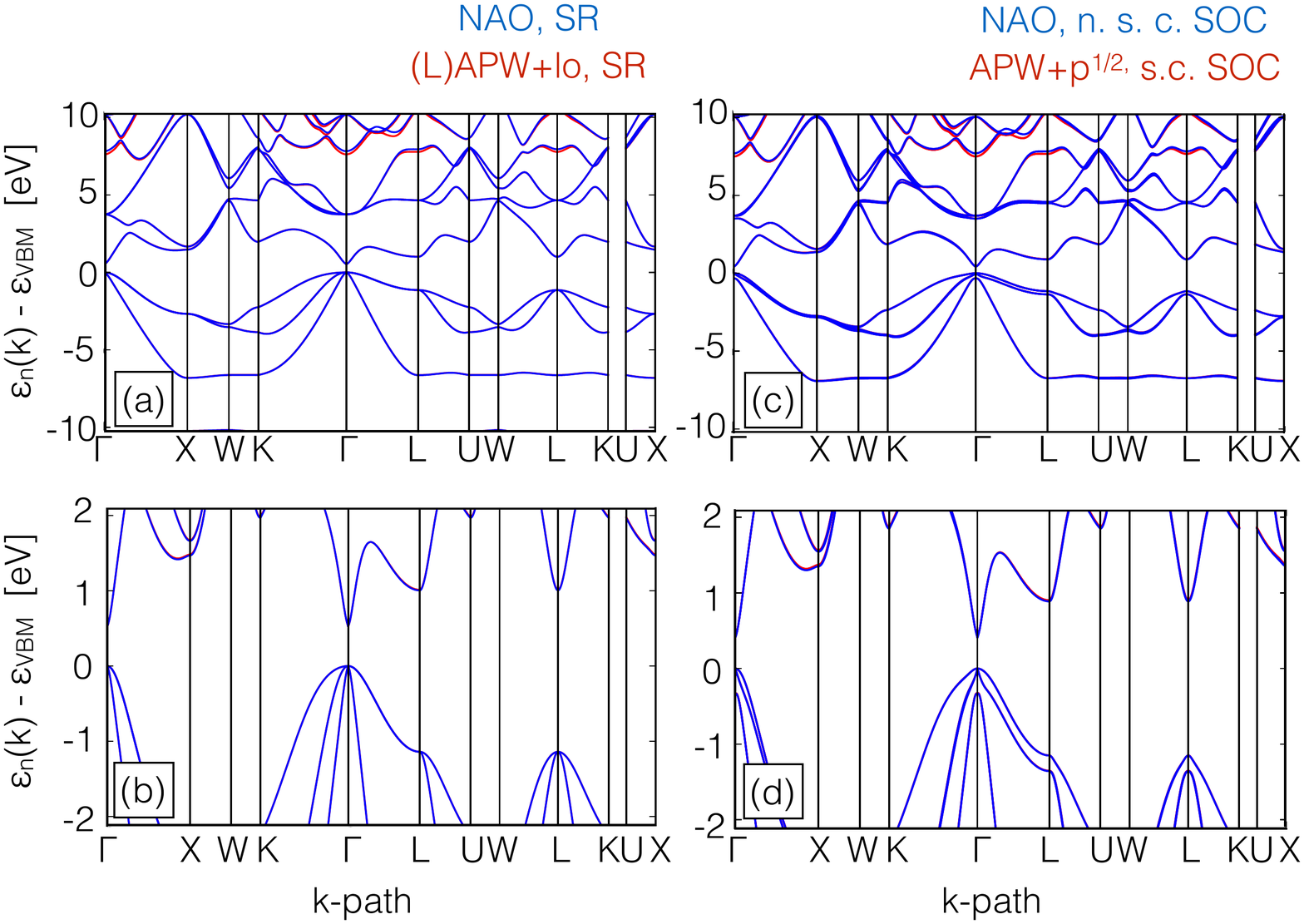}
  \caption{A comparison of PBE band structures of GaAs calculated using FHI-aims (tier 2) and WIEN2k ($RK_{max} = 10$) on (a) SR and (c) SOC levels of theory.  Close-ups on the band edges for SR and SOC levels of theory are shown in (b) and (d) respectively.  The experimental room-temperature lattice parameter of 5.6532 \AA~is  used.  Slight differences can be seen between NAO n.s.c. SOC and APW+p$^{1/2}$ s.c. SOC at X and in the $\Gamma \rightarrow$ L direction.  Band structures calculated using APW s.c. and APW+p$^{1/2}$ s.c. SOC are visually indistinguishable.}
\label{fig:GaAsBandStructure}
\end{figure*}

We next exemplify the differences that arise for scalar-relativistic and spin-orbit coupling treatments, as well as from different basis sets choices.  We will focus on the calculated band structure of the semiconductor GaAs\cite{Pearson1985} (cubic zincblende structure, experimental lattice paramemeter $a$=5.6532 \AA), but we will briefly assess other compound semiconductors for which VBM spin-orbit splittings been reported in the literature.  Figure \ref{fig:GaAsBandStructure}(a) overlays scalar-relativistic band structures at the level of DFT-PBE, calculated using (L)APW+lo (red) and a tier 2 NAO basis set (blue). Figure \ref{fig:GaAsBandStructure}(b) shows the segment around the band gap on a five times larger scale. In the physically important range of the valence bands and low-lying conduction bands (up to 5~eV above the conduction band minimum [CBM]) both sets are visually practically indistinguishable. Some visible differences arise only in the higher-energy bands (more than 7~eV above the CBM). These differences are a consequence of the more limited size of the NAO basis set (designed for accurate representations of occupied orbitals and the resulting density in DFT), compared to the overall larger (L)APW+lo basis set. For any scenarios requiring quantitatively more precise higher-lying states, further increasing the size of the NAO basis set used would be an adequate approach.

Similarly, Figure \ref{fig:GaAsBandStructure}(c) shows overlayed DFT-PBE band structures for GaAs including SOC, overlaying NAO n.s.c. band structures with APW+p$^{1/2}$ s.c. band structures. Excellent visual agreement for the valence and low-lying conduction bands shows that the n.s.c. SOC treatment based on atomic ZORA yields results that are essentially identical to the more sophisticated APW+p$^{1/2}$ s.c. treatment.  In particular, the hallmark SOC-related features in the GaAs band structure are precisely reproduced. The fundamental gap of GaAs decreases by approximately 100 meV when applying SOC.  This decrease arises from the splitting of the SR $\Gamma_{25'v}$ state into SOC $\Gamma_{8v}$ and $\Gamma_{7v}$ states, where $\Gamma_{8v}$ is the new VBM.  This splitting drives up the VBM (here defined to be the zero of energy) and consequently reduces the \emph{relative} energy of the conduction band $\Gamma_{6c}$, which itself is negligibly affected by SOC due to its s-like nature.

\begin{table}
	\centering
	\begin{tabular}{ |c||c|c||c|c|c||c|c| }
		\hline
                & NAO     & APW      & NAO    & APW & APW+p$^{1/2}$  & NAO     & Exp. \\
                & PBE     & PBE      & PBE    & PBE & PBE      & HSE06    & \\
                & SR      & SR       & SOC    & SOC & SOC      & SOC      & \\
                &         &        & n.s.c. & s.c. & s.c.          & n.s.c.  &      \\ 
        \hline
        \hline
        $\Gamma$\subscript{V}    & 0.000  &   0.000 &  0.000  & 0.000 &  0.000 &  0.000 &  0.00 \\
        $\Gamma$\subscript{SO}   & N/A    &   N/A   & -0.340 & -0.329 & -0.332 & -0.323 & -0.34 \\
		L\subscript{V}           & -1.139 &  -1.139 & -1.150 & -1.149 & -1.148 & -1.125 & -1.30 \\
		X\subscript{V}	         & -2.674 &  -2.672 & -2.746 & -2.742 & -2.742 & -2.988 & -2.80 \\
		\hline
        \hline
		$\Gamma$\subscript{C}    & 0.527  &   0.526 &  0.418 &  0.417 &  0.416 &  1.210 &  1.52 \\
		L\subscript{C}           & 1.004  &   1.012 &  0.889 &  0.902 &  0.902 &  1.596 &  1.78 \\
		X\subscript{C}           & 1.467  &   1.481 &  1.353 &  1.371 &  1.371 &  1.964 &  2.00 \\
		\hline
	\end{tabular}
	\caption{Comparison of calculated single-particle energy levels of GaAs with experimental quasi-particle energy levels.  Values are presented in units of eV. The first column labels the associated high-symmetry k-point, with a subscript of ``V'' denoting the valence band and ``C' the conduction band.  $\Gamma$\subscript{SO} denotes the split-off valence band at $\Gamma$.  The experimental room-temperature lattice parameter\cite{Pearson1985} of 5.6532 \AA~is  used.  Experimental energy levels are taken from Adachi\cite{Adachi04}.}
\label{tab:GaAsLevels}
\end{table}

In Table  \ref{tab:GaAsLevels}, we provide quantitative values for the valence and conduction band edges at the SR NAO, SR (L)APW+lo, NAO n.s.c. and APW+p$^{1/2}$ s.c. levels of theory for DFT-PBE at three high-symmetry $k$-points.  We place this comparison in the perspective of the quantitatively more accurate HSE06 density functional and of experimentally obtained energy levels at the same $k$-points (taken from Ref.\cite{Adachi04}). For the SR energy levels, agreement within 14~meV is observed; similarly, agreement within 19~meV is observed for the two different treatments of SOC. As is well known, the DFT-PBE level of theory itself shows key differences to experiment, which are partially alleviated by the more expensive HSE06 treatment. 

\begin{table}
	\centering
    \begin{threeparttable}
	\begin{tabular}{ |c||c|c|c||c| }
\hline
Structure & NAO    & APW  & APW+p$^{1/2}$ & Exp. (Ref.) \\
          & PBE    & PBE  & PBE           &             \\
          & SOC    & SOC  & SOC           &             \\
          & n.s.c. & s.c. & s.c.          &             \\
\hline
\hline
C\textsuperscript{a}    & 0.01 & 0.01 & 0.01 & 0.01 (Ref.\cite{OMadelung91}) \\
\hline
Si         & 0.05 & 0.05 & 0.05 & 0.04 (Ref.\cite{OMadelung91}) \\
ZnS\textsuperscript{b}  & 0.03 & 0.03 & 0.03 & 0.09 (Ref.\cite{OMadelung91}) \\
\hline
GaN\textsuperscript{c}  & 0.01 & 0.01 & 0.01 & 0.02 (Ref.\cite{Adachi04}) \\
GaP        & 0.08 & 0.08 & 0.08 & 0.08-0.1 (Ref.\cite{Adachi04}) \\
AlAs       & 0.30 & 0.30 & 0.30 & 0.28-0.33 (Ref.\cite{Adachi04}) \\
GaAs       & 0.34 & 0.33 & 0.33 & 0.34 (Ref.\cite{Adachi04}) \\
ZnSe       & 0.39 & 0.38 & 0.38 & 0.40 (Ref.\cite{OMadelung91}) \\
CdS\textsuperscript{b}  & 0.07 & 0.07 & 0.07 & 0.06 (Ref.\cite{OMadelung91}) \\
CdSe\textsuperscript{b} & 0.39 & 0.37 & 0.38 & 0.42 (Ref.\cite{OMadelung91}) \\
\hline 
InP        & 0.10 & 0.09 & 0.10 & 0.10-0.12 (Ref.\cite{Adachi04}) \\
AlSb       & 0.70 & 0.64 & 0.66 & 0.67 (Ref.\cite{Adachi04}) \\
ZnTe       & 0.93 & 0.83 & 0.89 & 0.91 (Ref.\cite{OMadelung91}) \\
CdTe       & 0.90 & 0.80 & 0.86 & 0.80 (Ref.\cite{OMadelung91}) \\
HgSe       & 0.25 & 0.22 & 0.23 & 0.38-0.40 (Refs.\cite{MDobrowolska80,RGalazka80,AMycielski82}) \\
HgTe       & 0.83 & 0.71 & 0.78 & 0.89-0.93 (Refs.\cite{NOrlowski00,CJanowitz01}) \\
\hline
	\end{tabular}
    \begin{tablenotes}
      \small
      \item\textsuperscript{a}Diamond structure.
      \item\textsuperscript{b}Wurtzite structure.
      \item\textsuperscript{c}Cubic zincblende structure.
    \end{tablenotes}
    \end{threeparttable}
	\caption{Comparison of PBE-calculated spin-orbit splittings to experimental values for the valence band of select compound semiconductors.  Values are presented in units of eV.  Benchmark Settings were used.}
\label{tab:ExpVBMSplits}
\end{table}

We end this section with a brief comparison of PBE-calculated spin-orbit splittings to experimental values for the valence bands of select compound semiconductors (Table \ref{tab:ExpVBMSplits}).  General agreement between calculated and experimental values to within 50 meV is observed for lighter materials, consistent with earlier validation work performed by Peralta \textit{et al.}\cite{JPeralta06} and Carrier and Wei.\cite{PCarrier04}.  For the heaviest materials considered (CdTe, HgSe, HgTe), deviations on the order of 200 meV are observed.  Notably, deviations between calculated and experimental values for HgSe are similar in magnitude to the calculated spin-orbit splittings themselves.  Similar disagreement was observed for mercury chalcogenides on the LDA level by Sakuma \textit{et al.}\cite{RSakuma11}, who attributed the deviation to a lack of many-body renormalization effects on the KS level of theory.  We will return to the question of choice of level of theory (in the context of density functionals) later in this paper.

\subsection{Non-Self-Consistent vs. Self-Consistent Treatment of Spin-Orbit Coupling: Trends Across 103 Materials}

\subsubsection{Quantifying Band Structure Differences: ``Band Delta''}
\label{sec:band_delta}

We first define a simple, quantitative metric for the difference between two calculated band structures, called ``band delta'' or $\Delta_{band}$ in the following. $\Delta_{band}$ is analogous to a root mean square deviation, defined on the energy levels of two energy band structures \{$\epsilon_{1,n}[\bm{k}_{i}]$\} and \{$\epsilon_{2,n}[\bm{k}_{i}]$\} within a given energy window [$-\epsilon_{l}$,$\epsilon_{u}$]:
\begin{equation}
	\Delta_{band}[-\epsilon_{l},\epsilon_{u}] = \sqrt{\frac{1}{N_{E}}\sum_{i=1}^{N_{k}}\sum_{\substack{\epsilon_{n,1}\le \epsilon_{u}\\\epsilon_{n,2}\le \epsilon_{u}\\\epsilon_{n,1}\ge -\epsilon_{l}\\\epsilon_{n,2}\ge -\epsilon_{l}}}(\epsilon_{n,1}[\bm{k}_{i}]-\epsilon_{n,2}[\bm{k}_{i}])^{2}} \, .
    \label{eq:BandDelta}
\end{equation}
$N_{k}$ is the number of unique k-points calculated along the k-path, $N_{E}$ is the total number of energy eigenvalues across all $\bm{k}_{i}$ that both band structures predict to lie within the energy window [$-\epsilon_{l}$,$\epsilon_{u}$], and $\epsilon_{1,n}[\bm{k}_{i}]$ and $\epsilon_{2,n}[\bm{k}_{i}]$ are the energy eigenvalues for the two band structures being compared at the k-point $\bm{k}_{i}$. For example, the SR valence bands of GaAs shown in Figure \ref{fig:GaAsBandStructure} have a $\Delta_{band}$[VBM-10 eV, VBM] value of 4 meV.
\begin{figure*}
    \includegraphics[trim={4cm 4cm 4cm 3.9cm},clip,width=\textwidth]{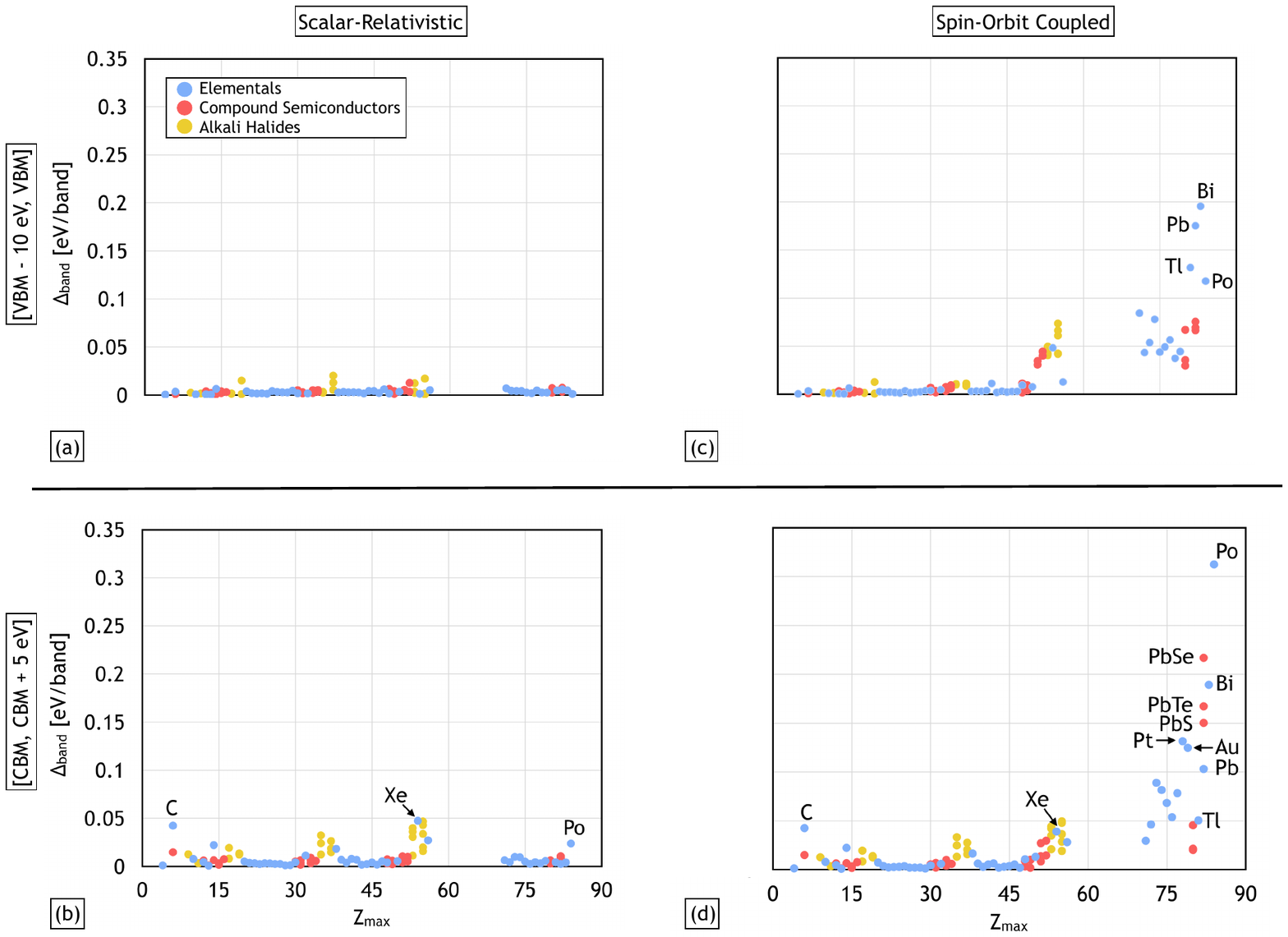}
    \caption{$\Delta_{band}$ on the benchmark set of 103 materials calculated with the PBE functional (a-b) for SR calculations comparing NAO and (L)APW+lo basis sets and (c-d) for SOC calculations comparing NAO n.s.c. and APW$+p^{1/2}$ s.c. handlings of SOC.  Outliers have been labeled.
}
    \label{fig:RMSValence}
\end{figure*}

\subsubsection{``Band Delta'' Analysis of Valence Bands of 103 Compounds}

  Figure \ref{fig:RMSValence}a and \ref{fig:RMSValence}b show $\Delta_{band}$ values between SR band structures for 103 elemental and multi-atomic compounds (Table \ref{tab:Materials}), calculated using NAO and (L)APW+lo basis sets, respectively. Since the purpose of this paper is to highlight differences of different SOC treatments, we focus on the range of valence bands [VBM-10 eV, VBM] and low-lying conduction bands [CBM, CBM + 5 eV], i.e., the energy windows for which both basis set types are expected to give the most accurate answers.  The $\Delta_{band}$ values in Fig.~\ref{fig:RMSValence} are ordered according to the maximum atomic number Z\subscript{max} within a given material, i.e., the principal quantity with which relativistic effects are expected to increase.

In brief, at the SR level of theory, we find excellent agreement between NAO and (L)APW+lo basis sets for the valence and low-lying conduction bands of all compounds. A maximum $\Delta_{band}$ value of 20 meV in the valence bands and 47 meV in the low-lying conduction bands is observed for alkali halides containing three specific elements:  the large alkali atoms K ($Z$=19), Rb ($Z$=37), and Cs ($Z$=55).  The reason why these specific elements stand out is unclear and could reside \emph{either} on the NAO side \emph{or} on the (L)APW+lo side, or both. (Regarding NAOs, we note that no special behavior or uncertainties regarding alkali atoms were observed during the construction of the NAO basis sets as reported in Ref. \cite{Blum09}.)  However, even the maximum $\Delta_{band}$ for these three outliers is so minor that it does not affect the conclusions regarding SOC, which is the main purpose of this work.

In Figures \ref{fig:RMSValence}c and \ref{fig:RMSValence}d, an analogous comparison of $\Delta_{band}$ is reported, but this time for NAO n.s.c. SOC vs. APW+p$^{1/2}$ s.c. SOC band structures. Remarkably, for materials with $Z\subscript{max} \le 50$ (Sn), $\Delta_{band}$ for SOC band structures are similar to those seen in SR band structures, i.e., extremely low.  The ``outlier materials'' containing K, Rb, and Cs also agree somewhat better for valence bands between both methods. This suggests that the changed basis set on the (L)APW+lo side may have some accidental beneficial effect. (The NAO basis set remains unchanged compared to the SR case.) As shown in Figure~\ref{fig:GaAsBandStructure} and Table~\ref{tab:GaAsLevels}, the effects of SOC in the $Z\subscript{max} \le 50$ range can already be appreciable. However, they appear to be captured essentially exactly by the n.s.c. approach to SOC. This justifies the use of the computationally relatively cheap post-processing of a s.c. SR calculation even for high-accuracy band structures at least up to $Z\subscript{max} \approx 50$.

From Z\subscript{max} = 51 (Sb) to Z\subscript{max} = 80 (Hg), Fig.~\ref{fig:RMSValence}b shows elevated but consistent $\Delta_{band}$ values , up to $\Delta_{band}$=84 meV (Z\subscript{max} = 71, Lu) for valence bands and 131 meV (Z\subscript{max} = 78, Pt) for low-lying conduction bands. Most $\Delta_{band}$ lie between 30 meV and 70 meV. This implies that n.s.c. SOC should safely serve to capture any qualitatively relevant band structure effects even in this range. The associated uncertainty is well below the overall uncertainty implied, e.g., by the use of DFT itself, and potentially other approximations inherent in the Born-Oppenheimer treatment of materials in most standard computations.

For the heaviest materials in the benchmark set (Z\subscript{max} = 81 onwards), $\Delta_{band}$ as large as 196 meV (Z\subscript{max} = 81, Bi) for valence bands and 312 meV (Z\subscript{max} = 83, Po) for low-lying conduction bands are observed. These elements feature open 6$p$ valence shells, i.e., the shell that is most affected by SOC on a qualitative level (see Fig.~\ref{fig:Hg_Orbitals} and the associated discussion).

The quantitative band structure deviations associated with 6$p$ elements are thus significant (a fact well known in the community\cite{JCBoettger00,JKunes01,PLarson03,PCarrier04}).  However, as we show specifically for spin-orbit splittings below, even in this range the \emph{relative} accuracy of n.s.c. SOC (compared to the overall magnitude of SOC effects) is still within 11\%, justifying the use of n.s.c. SOC treatments for qualitative analyses of band structures even for very heavy elements.  
\subsubsection{Spin-Orbit Splittings}

\begin{figure}
    \includegraphics[trim={9.75cm 1.5cm 9.5cm 1cm},clip,width=0.5\textwidth]{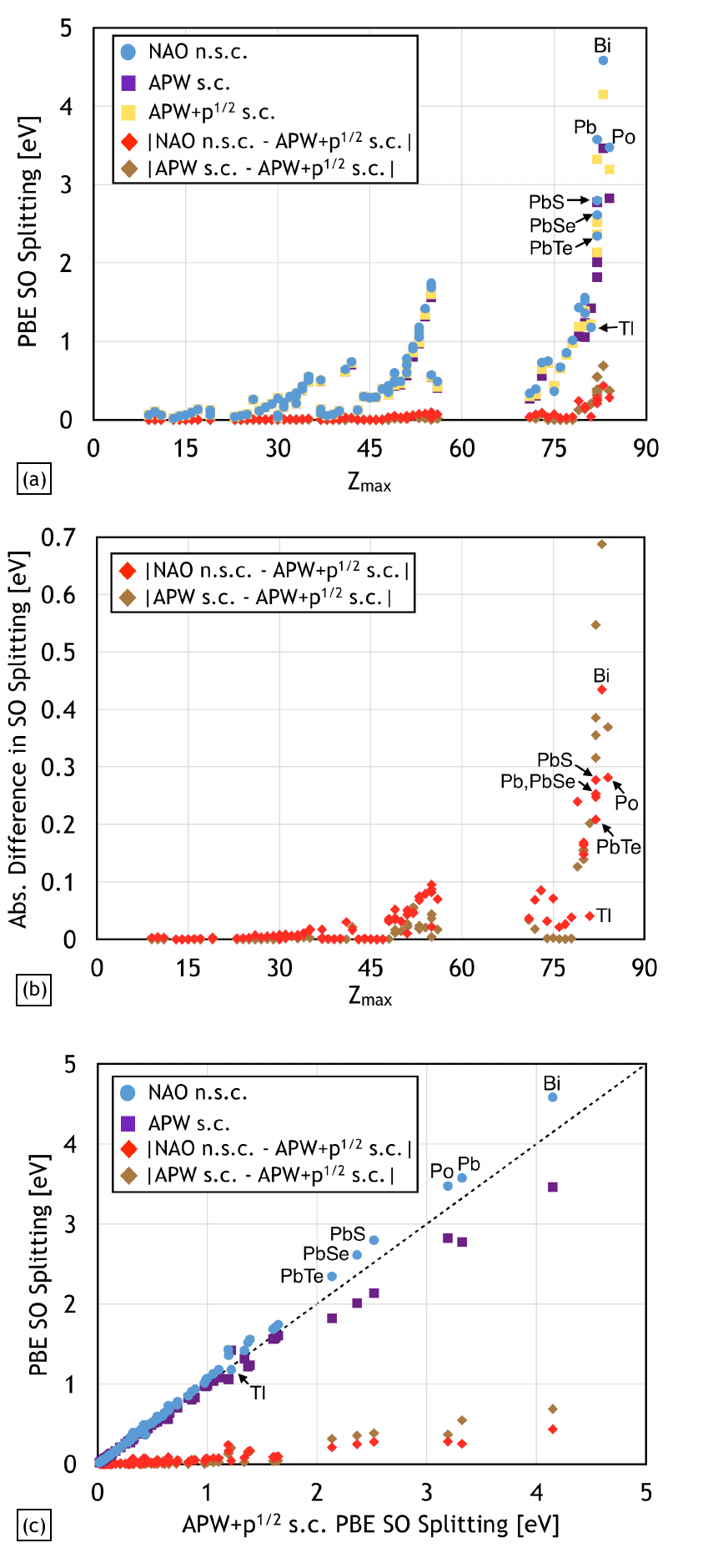}
    \caption{NAO n.s.c., APW s.c., and APW+p$^{1/2}$ s.c. spin-orbit splittings for the benchmark set of 103 materials calculated using the PBE functional.  In subfigure (c), the y=x line corresponds to perfect agreement between APW+p$^{1/2}$ s.c. SOC and the other two levels of theory.}
  \label{fig:SOSplittingsPBE}
\end{figure}

Figure \ref{fig:SOSplittingsPBE} shows spin-orbit splittings $\Delta_{SOC}$ at specific $k$ points,
calculated using NAO n.s.c., APW s.c., and APW+p$^{1/2}$ s.c. SOC and the PBE functional.  Benchmark Settings are used.  For each material in our band structure benchmark set, we select the largest unambiguous spin-orbit splitting of valence and conduction states for inclusion in this study.  Materials for which spin-orbit splittings could not be unambiguously identified by visual inspection of band structures were omitted.  The spin-orbit splitting chosen for each material are provided in Section 5 of the SM.  

Figure \ref{fig:SOSplittingsPBE}a illustrates the expected strong dependence of the magnitudes of spin-orbit splittings on $Z_{max}$. In addition to the values of the splittings, the difference between the calculated NAO n.s.c. and APW+p$^{1/2}$ s.c. splittings is also shown. Interestingly, for the selected splittings, the magnitude of the difference remains very small compared to the overall spin-orbit splittings, indicating relative deviations of the n.s.c. approach are within 11$\%$ for the heaviest tested materials in this plot (Tl, Pb, Pb chalcogenides, Bi, and Po).

For a more quantitative comparison, we turn to Fig.~\ref{fig:SOSplittingsPBE}b, which shows the difference between the expected highest-accuracy APW+p$^{1/2}$ s.c. spin-orbit splittings and two different approximate approaches: NAO n.s.c. spin-orbit splittings and APW s.c. spin-orbit splittings, respectively, again as a function of $Z\subscript{max}$. Qualitatively, the observed deviations for these specific spin-orbit splittings are well in line with the broader comparison of $\Delta_{band}$ values in Fig.~\ref{fig:RMSValence}.  For materials exclusively containing elements lighter than 6p, we find a maximum deviation of 156 meV (HgS) between NAO n.s.c. and APW+p$^{1/2}$ s.c. spin-orbit splittings.  This deviation is within the margin of error expected due to usage of the (g)KS level of theory, as previously shown in Table \ref{tab:ExpVBMSplits}.

Interestingly, both NAO n.s.c. and APW s.c. spin-orbit splittings differ noticeably from APW+p$^{1/2}$ s.c. spin-orbit splittings for materials containing 6p elements, suggesting that the dominant contribution is the purely SR nature of the basis set considered and not the approximation of non-self-consistency used in the calculation.  The deviations of the NAO n.s.c. splittings are actually \emph{smaller} than that of the APW s.c. splittings. Maximum differences of 688 meV (APW s.c. SOC) and 435 meV (NAO n.s.c. SOC) from APW+p$^{1/2}$ s.c. SOC values were observed for Bi (Z\subscript{max} = 83.)  Although this observation is difficult to generalize, it could be that the location of the atomic ZORA $p$ functions between the $p^{1/2}$ and $p^{3/2}$ radial functions (Figure~\ref{fig:Hg_Orbitals}) renders the atomic ZORA $p$ function a slightly better starting point for the second-variational treatment without explicit $p^{1/2}$ radial functions.

We replot the splittings of Fig.~\ref{fig:SOSplittingsPBE}a in Fig.~\ref{fig:SOSplittingsPBE}c as a function of the APW+p$^{1/2}$ s.c. splittings. This reveals an interesting trend, namely that NAO n.s.c. SOC consistently predicts larger values and APW s.c. SOC generally predicts smaller values for the splittings than the reference approach, i.e., APW+p$^{1/2}$ s.c. SOC. This trend may be related to the difference of the underlying SR core orbitals of both approaches.  

\begin{table}
	\centering
    \begin{threeparttable}
	\begin{tabular}{ |c|c||c|c|c||c| }
		\hline
        Material & VB         & NAO     & APW   & APW+p$^{1/2}$ & Exp. \\
                 & Spin-Orbit & SOC     & SOC   & SOC           &      \\
                 & Splitting  & n.s.c.  & s.c.  & s.c.          &      \\ 
        \hline
        PbS      & $\Gamma^{-}_{8}-\Gamma^{-}_{6}$ & 0.37    & 0.31  & 0.36          & 0.3\textsuperscript{a} \\
        PbS      & $X^{-}_{7}-X^{-}_{6}$           & 0.33    & 0.28  & 0.32          & 0.2\textsuperscript{a} \\
        PbSe     & $\Gamma^{-}_{8}-\Gamma^{-}_{6}$ & 0.69    & 0.60  & 0.67          & 0.6\textsuperscript{a}, 0.75\textsuperscript{b} \\
        PbSe     &  $X^{-}_{7}-X^{-}_{6}$          & 0.51    & 0.43  & 0.49          & 0.5\textsuperscript{a}, 0.55\textsuperscript{b} \\
        PbTe     & $\Gamma^{-}_{8}-\Gamma^{-}_{6}$ & 1.18    & 1.00  & 1.12          & 1.15\textsuperscript{a}, 1.10\textsuperscript{b} \\
        PbTe     &  $X^{-}_{7}-X^{-}_{6}$          & 0.77    & 0.63  & 0.72          & 0.9\textsuperscript{a} , 1.10\textsuperscript{b} \\
        \hline
	\end{tabular}
    \begin{tablenotes}
      \small
      \item\textsuperscript{a}Taken from Ref.\cite{TGrandke78}
      \item\textsuperscript{b}Taken from Ref.\cite{VHinkel89}
    \end{tablenotes}
    \end{threeparttable}
	\caption{Comparison of PBE-calculated valence band spin-orbit splittings for lead chalogenides.  Values are presented in units of eV.  Benchmark Settings were used.}
\label{tab:PbChalogenides}
\end{table}

The spin-orbit splittings reported in Figure \ref{fig:SOSplittingsPBE} were chosen by the criterion of being the largest spin-orbit splitting observed in the band structure for a material within a [-10 eV, 10 eV] energy range around the VBM/Fermi level.  Band structures contain multiple spin-orbit splittings in this energy range, and the largest spin-orbit splitting cleanly observable in a computed band structure may be difficult to observe in experimental spectra (e.g. it may lie in the middle of the conduction band).  This was observed for the lead chalcogenides, for which the spin-orbit splittings reported in Figure \ref{fig:SOSplittingsPBE} lie within the conduction band at $\Gamma$.

We provide the calculated valence band spin-orbit splittings for the lead chalcogenides alongside experimental values in Table \ref{tab:PbChalogenides}.  Notably smaller maximum deviations were observed for VB splittings (60 meV for NAO n.s.c., 120 meV for APW s.c.) compared to the maximum deviations in the CB splitting at $\Gamma$ for lead chalcogenides reported in Figure \ref{fig:SOSplittingsPBE} (280 meV for NAO n.s.c., 390 meV for APW s.c.).  The improved agreement is consistent with the trends reported in this paper, as the VB splittings arise primarily from the p orbitals of the lighter chalcogenides (though with some contribution from Pb p orbitals\cite{TGrandke78}) whereas the CB splitting at $\Gamma$ are dominated by the p orbitals of heavier lead.  Calculated spin-orbit splittings in Table \ref{tab:PbChalogenides} are generally in agreement with experiment, with the exception of the $X$ splitting for PbTe where calculated values underestimate the splitting on the order of 400 meV.  Similar results were observed by Svane \textit{et al.}\cite{ASvane10} at the level of quasiparticle self-consistent \textit{GW} many-body theory, also employing a post-processed non-self-consistent treatment of SOC.

\begin{figure*}
  \includegraphics[trim={0cm 0cm 0cm 0cm},clip,width=\textwidth]{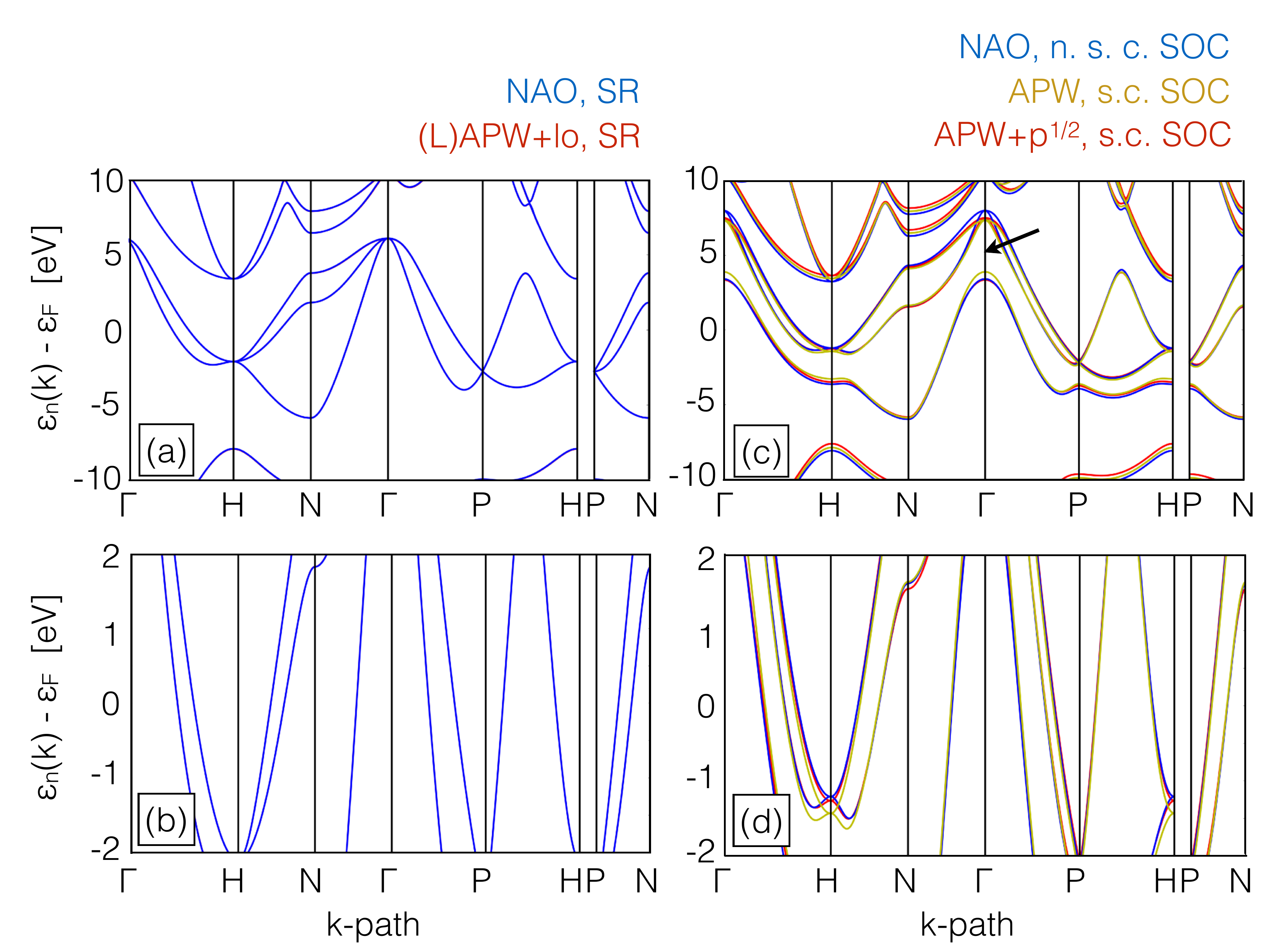}
  \caption{A comparison of PBE band structures of BCC Bi calculated using FHI-aims and WIEN2k on (a) SR and (c) SOC levels of theory.  Close-ups near the Fermi level for SR and SOC levels of theory are shown in (b) and (d) respectively. The midpoint of the spin-orbit splitting reported in the text is marked by an arrow in subfigure (c).  Benchmark Settings and the experimental room-temperature lattice parameter\cite{Pearson1985} of 3.800 \AA~were used.  Scalar-relativistic band structures calculated using NAO and (L)APW+lo basis sets are visually indistinguishable.}
\label{fig:BiBandStructure}
\end{figure*}

Returning to the spin-orbit splittings reported in Figure \ref{fig:SOSplittingsPBE}, we end this section by putting the largest observed quantitative deviations into proper context with a comparison of the computed band structures for Bi (Figure \ref{fig:BiBandStructure}).  The reported spin-orbit splitting occurs at the lowest-lying conduction band at $\Gamma$ and is marked by an arrow in subfigure (c).  The SOC-perturbed band structures on the three levels of theory, though visually distinguishable, all predict the same qualitative behavior.  In the region near the Fermi level (subfigures b and d) critical for modeling electronic transport properties, the three levels of theory are vertically shifted relative to one another, but this translate to small lateral shifts in the Fermi surface in practice.  The lingering quantitative deviations induced by usage of NAO n.s.c. versus APW+p$^{1/2}$ s.c. SOC are much smaller than the qualitative (and quantitative) improvement relative to the original SR band structure.  Overall, Fig.~\ref{fig:SOSplittingsPBE}c and \ref{fig:BiBandStructure} together visually reaffirm the impression that the NAO n.s.c. SOC approximation to SOC still captures SOC effects on the band structure correctly even for the heaviest elements investigated here.  

\subsubsection{PBE versus HSE06}

  We next investigate the dependence of calculated SOC effects on the underlying density functional, using the PBE and HSE06 functionals as examples.  The NAO n.s.c. SOC approach with Tight Production Settings is used.  We consider spin-orbit splittings as well as changes of the fundamental gap changes due to SOC,
\begin{equation}
  \Delta E_{g} = E^{SOC}_{g} - E^{SR}_{g} \, .
\end{equation}
Here $E^{SR}_{g}$ is the scalar-relativistic fundamental gap and $E^{SOC}_{g}$ is the spin-orbit-coupled fundamental gap.  The fundamental gaps and spin-orbit splittings calculated using Tight Production Settings for gapped materials are listed in Section 6 of the SM.  We omit Co and Fe from the comparison of spin-orbit splittings at Tight Production Settings due to difficulty converging the electronic structures on the HSE06 level of theory.

\begin{figure}
    \includegraphics[trim={9.5cm 1.5cm 9.5cm 1cm},clip,width=0.5\textwidth]
{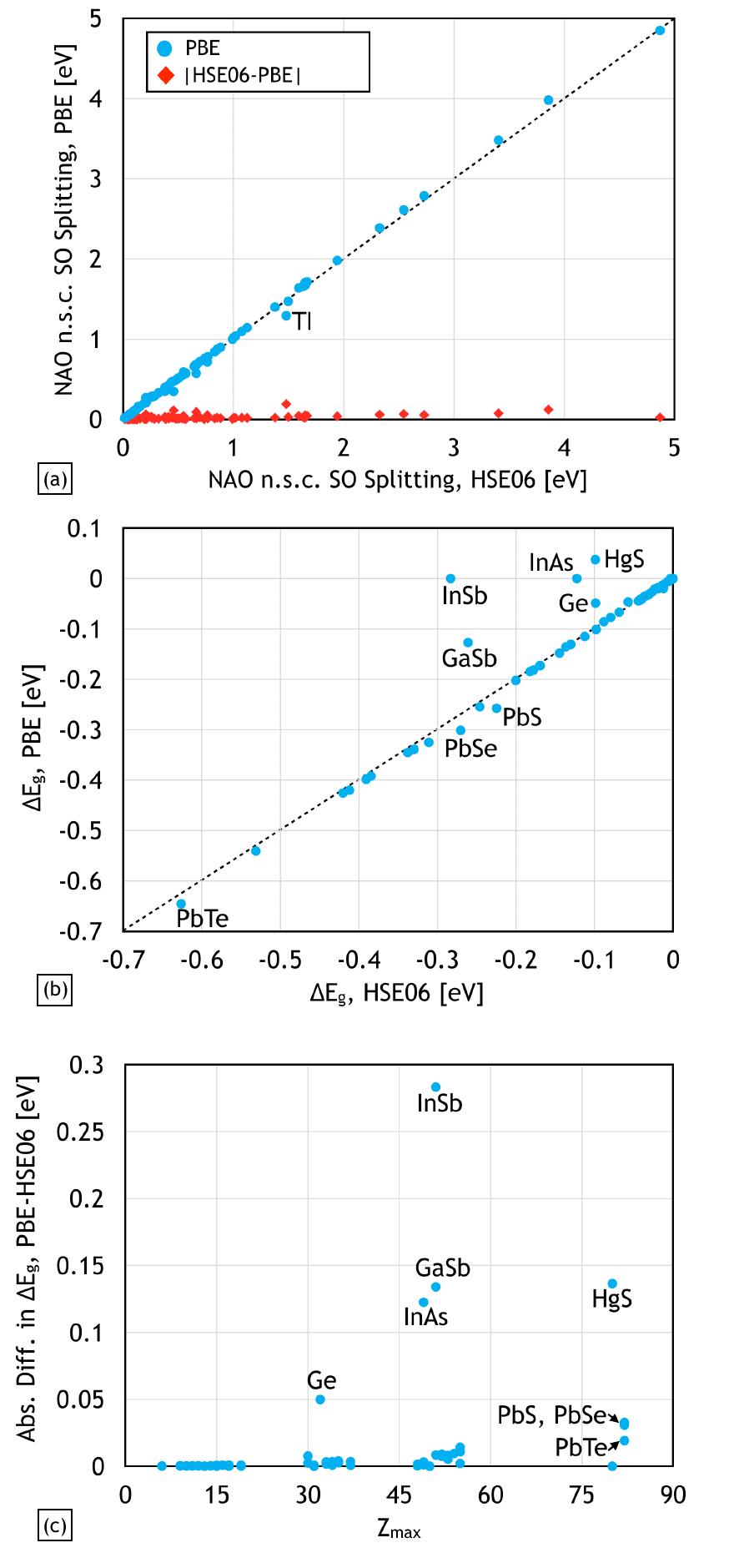}
    \caption{A comparison between the PBE and HSE06 functionals using NAO n.s.c. SOC for (a) spin-orbit splittings and (b) SOC-induced fundamental gap changes.  Figure (c) shows the absolute deviation in SOC-induced fundamental gap changes.}
  \label{fig:FunctionalDependence}
\end{figure}

Figure \ref{fig:FunctionalDependence}a shows spin-orbit splittings calculated with the PBE and HSE06 functionals and supports the assertion in the literature\cite{JPeralta06} that different functionals yield similar spin-orbit splittings.  Most differences lie below 60 meV.  A notable exception is Tl, with an absolute difference of 189 meV.  We note that the spin-orbit splitting chosen for Tl occurs for a high-lying state, and this deviation is likely due to a combination of the basis set size used for these calculations and the open-shell 6p orbital.

We next turn to SOC-induced changes of the band gap, $\Delta E_{g}$, in Figure \ref{fig:FunctionalDependence}b.  Only gapped materials are included in the $\Delta E_{g}$ comparison.  We note that SOC consistently \emph{decreases} the fundamental gap relative to the SR fundamental gap ($\Delta E_{g} \le 0$).  

Regarding the dependence on the density functional used, it is striking that there is no significant functional dependence of the SOC-induced gap \emph{change}, except for a few materials that show surprisingly stark differences between the PBE and HSE06 functional (Figure \ref{fig:FunctionalDependence}c).  We discuss these materials in more detail in the next subsection.	

\subsubsection{The Outliers:  Ge, InAs, GaSb, InSb, HgS}
 
The five ``outlier'' materials (Ge, InAs, GaSb, InSb, HgS) in Fig.~\ref{fig:FunctionalDependence}b are either predicted to be zero-band-gap semiconductors (InAs, InSb, HgS) or have negligible fundamental gaps of 0.13 eV or less (Ge, GaSb) in SR DFT-PBE. In contrast, in SR HSE06, they have gaps of 0.35 eV or more.

\begin{figure}
  \includegraphics[trim={2cm 3cm 0cm 2.5cm },width=0.5\textwidth]{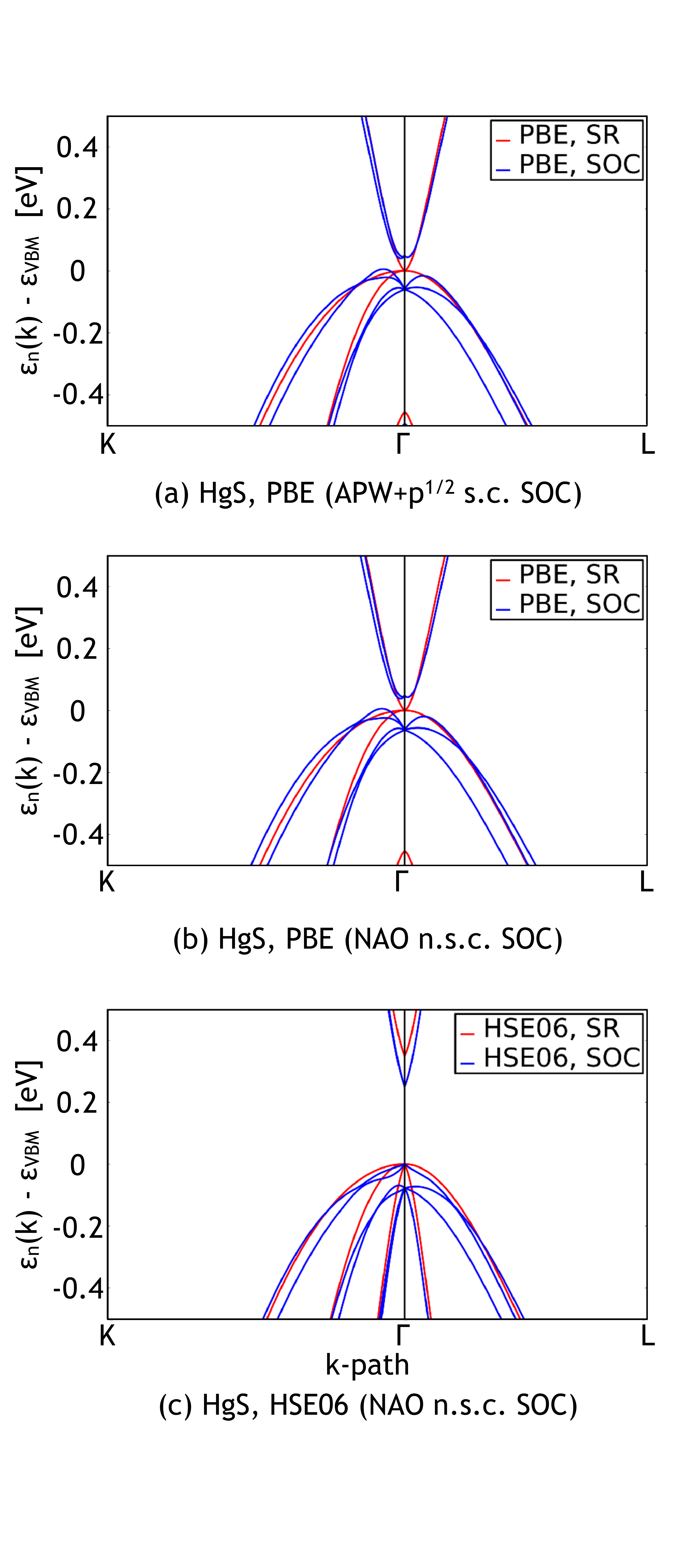}
    \vspace{-1.5cm}
    \caption{Scalar-relativistic and spin-orbit-coupled band structures near band edges of zincblende HgS calculated at Tight Production Settings using (b) the PBE functional and (c) the HSE06 functional.  The band structures calculated using the PBE functional and APW+p$^{1/2}$ s.c. SOC at Benchmark Settings (subfigure (a)) are also shown for comparison.}
  \label{fig:HgS}
\end{figure}

To pinpoint the origin of this behavior, in Figure~\ref{fig:HgS} we consider the example of HgS, considered here in the (cubic) zincblende $\beta$ phase.  The PBE functional predicts SR HgS to be a zero-band-gap semiconductor at a lattice parameter of 5.874 \AA.  SOC intermixes the valence and conduction bands at $\Gamma$ (Figure \ref{fig:HgS}b), opening an indirect fundamental gap of 38 meV.  A similar behavior is observed for Wien2k and APW+p$^{1/2}$ SOC at Benchmark Settings  (Figure \ref{fig:HgS}a).  The HSE06 functional predicts a separation of the valence and conduction band on the SR level of theory, and application of SOC preserves the direct nature of the fundamental gap (Figure \ref{fig:HgS}c).

Analogous observations were made by Aguilera \textit{et al.}\cite{IAguilera15},   Svane \textit{et al.}\cite{ASvane11}, and Sakuma \textit{et al.}\cite{RSakuma11} in the context of $GW$ quasiparticle energy calculations.  For mercury chalcogenides\cite{ASvane11,RSakuma11} and bismuth,\cite{IAguilera15} (g)KS-DFT predicts qualitatively incorrect features for SOC band structures. It is necessary to include SOC self-consistently within the GW method for these materials to achieve qualitatively accurate results\cite{FAryasetiawan08,FAryasetiawan09,RSakuma11}.  

\section{Conclusions}

  In this paper, we established a band structure benchmark set and systematically examined the dependence of SOC-related phenomena on basis sets (NAO, (L)APW+lo, and (L)APW+lo including a p$^{1/2}$ orbital), handling of self-consistency within post-processed SOC (n.s.c. and s.c.), and functionals (PBE and HSE06) across 103 materials and two different, high-accuracy all-electron DFT codes (FHI-aims and Wien2k). At the scalar-relativistic level of theory, near-complete agreement between band structures calculated with the different basis sets and codes is found. Regarding SOC effects, we find quantitative agreement between all basis sets and SOC approaches used up to the 5p block of the periodic table.    We provide these benchmark-quality band structures to the community via the NOMAD Repository\cite{NOMAD} citable by  Ref.\cite{Huhn2017BandStructureBenchmarkWIEN2k_BenchmarkSettings} for WIEN2k-calculated band structures and Ref.\cite{Huhn2017BandStructureBenchmarkFHI-aims_BenchmarkSettings} for FHI-aims-calculated band structures.  We also provide HSE06-calculated band structures calculated by FHI-aims, as well as the associated PBE band structures, citable by Ref.\cite{Huhn2017BandStructureBenchmarkFHI-aims_TightProductionSettings}
  
  For materials containing heavier atoms, divergences between the methods using purely SR valence bands (NAO n.s.c. and APW s.c. SOC) and APW+p$^{1/2}$ n.s.c. SOC are observed. For the heaviest elements (6$p$ valence shell) spin-orbit splittings calculated by APW s.c. SOC deviate from the APW+p$^{1/2}$ reference by as much as 0.69~eV. For the NAO n.s.c. SOC the discrepancy to APW+p$^{1/2}$ in the 6$p$ shell is much smaller.  We thus find qualitative agreement in PBE-calculated band structures predicted by the APW+p$^{1/2}$ s.c. SOC and NAO n.s.c. SOC treatments for all materials investigated here. For large-scale calculations where self-consistent SOC becomes computationally expensive, non-self-consistent SOC offers a convenient and qualitatively accurate method for approximating the necessary effects of spin-orbit coupling.

  We also compared spin-orbit splittings and fundamental gap changes due to SOC calculated using the semi-local PBE and hybrid HSE06 functional.  Close agreement in these quantities was observed between exchange-correlation functionals even for the heaviest materials.  However, this agreement requires qualitative agreement in the underlying SR band structures, as energy levels not properly gapped can intermix once SOC is applied.  The notion that different functionals yield similar SOC-calculated quantities thus comes with a caveat:  qualitatively accurate SR band structures are necessary to ensure that physically meaningful results emerge from second-variational SOC.

\section{Acknowledgements}

W.P.H. would like to thank Matthias Scheffler and the Fritz-Haber Institut Berlin for funding the early stages of this work via a ``fellowship to promote scientific cooperation with foreign countries.''  This work was partially supported by the LDRD Program of ORNL managed by UT-Battle, LLC, for the U.S. DOE and by the Oak Ridge Leadership Computing Facility, which is a DOE Office of Science User Facility supported under Contract DE-AC05-00OR22725.

\bibliographystyle{apsrev4-1}
\bibliography{wphbib}
\end{document}